\pdfoutput=1
\PassOptionsToPackage{unicode}{hyperref}
\PassOptionsToPackage{hyphens}{url}
\documentclass[12pt]{article}

\usepackage[T1]{fontenc}
\usepackage[utf8]{inputenc}
\usepackage{lmodern}
\usepackage{amsmath,amssymb,amsthm,mathtools,bm}
\usepackage[margin=1in]{geometry}
\usepackage{setspace}
\usepackage{booktabs,array,threeparttable}
\usepackage{enumitem}
\usepackage{framed}
\usepackage{graphicx}
\graphicspath{{figures/}}
\usepackage[authoryear,round]{natbib}
\usepackage{xurl}
\usepackage[colorlinks=true,linkcolor=blue,citecolor=blue,urlcolor=blue]{hyperref}
\hypersetup{
  pdftitle={Positive Weights Do Not Certify TWFE Inference: A Design Diagnostic under Treatment-Effect Heterogeneity},
  pdfkeywords={difference-in-differences, staggered adoption, treatment-effect heterogeneity, design diagnostics, cluster-robust inference}}

\newtheorem{theorem}{Theorem}
\newtheorem{proposition}{Proposition}

\theoremstyle{definition}
\newtheorem{assumption}{Assumption}
\newtheorem{definition}{Definition}
\theoremstyle{remark}

\DeclareMathOperator{\E}{\mathbb E}
\DeclareMathOperator{\Var}{Var}
\DeclareMathOperator{\Cov}{Cov}
\DeclareMathOperator{\tr}{tr}
\newcommand{\R}{\mathbb R}
\newcommand{\Pp}{\mathbb P}
\newcommand{\Dt}{\widetilde D}
\newcommand{\hbeta}{\widehat\beta}
\newcommand{\ATT}{\mathrm{ATT}}
\newcommand{\nw}{n_w}
\newcommand{\onevec}{\mathbf 1}
\newcommand{\eps}{\varepsilon}
\newcommand{\qhat}{\widehat q}
\newcommand{\chat}{\widehat c}
\newcommand{\Khat}{\widehat K}

\newcommand{\authorblock}{Stanis\l{}aw M. S. Halkiewicz\\Lule\aa{} University of Technology\\\texttt{stashal@o2.pl}}
  \newcommand{\softwareline}{The diagnostic is implemented in the open-source \textsf{PanelAdequacy} software \citep{PanelAdequacySoftware}.}
  \newcommand{\companionline}{The same inversion, at the same nominal size and tolerance, fixes the threshold in a companion paper on classical measurement error in saturated fixed-effect designs \citep{HalkiewiczEIV}; the tolerance convention is therefore common to the two diagnostics, although the pathologies, design objects and drifts differ.}

\newtheorem{lemma}{Lemma}
\DeclareMathOperator{\rank}{rank}
\newcommand{\hsigma}{\widehat\sigma}

\begin{document}
\onehalfspacing

\begin{center}
{\LARGE\bf Positive Weights Do Not Certify TWFE Inference: A Design Diagnostic under Treatment-Effect Heterogeneity}

\medskip
\authorblock 
\end{center}

\begin{abstract}
Conventional two-way fixed-effects (TWFE) inference need not be uniformly adequate even when all treatment weights are positive. Under a local treatment-effect drift, the statistic for the equally weighted treated-cell effect has worst-case mean shift $(c/q)\Gamma$, where $\Gamma=\sqrt{N_1}\|w-u\|_2$ is design-only and $q$ is the cluster-score noise scale. Projection matches the exposure to cohort, event-time, additive, and saturated group-time pilots; when residualized treatment is constant within treated group-time cells, $\Gamma_{\mathrm{gt}}=\Gamma$. A direct score estimator is feasible with fixed panel length as clusters grow. Because a debiased quadratic pilot is nonregular at zero, we use two regular one-sided procedures: a reverse-triangle multiplier lower bound to detect unsafe envelopes and covariance-aware noncentral-$\chi^2$ inversion to certify adequacy. At $5\%$ nominal size with a five-point tolerance, the shift threshold is $0.652$. In the Callaway--Sant'Anna minimum-wage panel, where every treated-cell weight is positive, the saturated $95\%$ lower shift bound is $0.793$, so conventional inference is not uniformly certified. A complementary Brazilian property-tax reanalysis with $2{,}840$ municipalities withholds certification in all four classes; its smallest lower bound is $1.040$. Under the maintained many-cluster approximation, castle-doctrine homicide passes for additive cohort-plus-event-time heterogeneity with a $95\%$ upper bound of $0.236$; its saturated covariance lacks one required rank. Divorce is inconclusive. These envelopes are not predictions of realized rejection. The diagnostic separates point correction, uniform exposure, and statistically warranted certification.
\end{abstract}


\newpage

\section{Introduction}\label{sec:intro}

Two-way fixed-effects (TWFE) regression remains one of the most familiar ways to study policies adopted in different places at different times. Its appeal is easy to understand. Unit effects absorb persistent differences across places, time effects absorb common shocks, and a single coefficient summarizes the estimated response. Combined with clustered standard errors, this specification has become a standard empirical language for policy evaluation.

The credibility revolution put research design and transparent identifying comparisons at the center of empirical economics \citep{AngristPischke2009}. In staggered policy settings, TWFE became a standard language for implementing that design. A second stage then clarified the estimand. \citet{dCdH} and \citet{GoodmanBacon} showed that under heterogeneous effects the TWFE coefficient is generally a nonuniformly weighted combination of treated-cell effects, sometimes with negative weights. \citet{CS}, \citet{SunAbraham}, and \citet{BorusyakEtAl} supplied estimators for interpretable group-time and event-time targets, and these methods are now part of routine applied practice.

Those two stages answer different questions. The first asks which comparisons identify an effect; the second asks which effect the estimator aggregates. Neither certifies that the reported test has its nominal size under the heterogeneity supported by the data. The credibility revolution addressed identification, and the recent heterogeneity literature repaired the estimand. The inference problem survived both.

The gap is systematic. Inference adequacy---whether a realized design can support the test actually reported---has usually been addressed problem by problem. Its symptoms recur across modern empirical work: \citet{Young2022} documents leverage concentration and distorted IV inference in published applications; \citet{KSS} develops leave-out variance methods for many-fixed-effect models; and few-cluster corrections remain a separate practical literature \citep{MNW}. The staggered-adoption weighting problem itself became explicit only after DiD had been standard practice for decades. These are different mechanisms, but they share a methodological lesson: credible identification and an interpretable estimand do not complete the inferential audit.

Applied work nevertheless relies on two reassuring shortcuts. Researchers may regard a conventional TWFE confidence interval as adequately protected when all treated-cell weights are positive or when the TWFE coefficient lies near a heterogeneity-robust aggregate. Neither observation answers the coverage question. Positive weights make the coefficient a convex average, but they do not make that average equal to the target. Coverage depends on the discrepancy relative to the standard error. A modest coefficient difference can be consequential in a precise panel, while visibly uneven weights can be harmless when heterogeneity is small relative to sampling noise.

This paper develops a diagnostic for that missing step. Given an adoption design, a target, and an empirically calibrated class of treatment-effect heterogeneity, it asks whether the conventional TWFE test can be certified to remain within a stated size tolerance uniformly over that class. $\Gamma$ is a worst-case envelope over the stated class, not a prediction of realized rejection. Failure to certify does not show that a published result is invalid; it means the design and calibration do not rule out profiles that violate the tolerance. A pass is the stronger conclusion: every profile in the stated class satisfies the bound. If the projected covariance lacks the rank required by that class, the procedure reports an inconclusive result for that structural reason rather than forcing a verdict.

The diagnostic separates three objects that are often blended together. The first is the design's exposure to heterogeneity: how far its treated-cell weights lie from the target weights, in the directions allowed by the chosen class. The second is the amount of heterogeneity supported by the outcome data. The third is the cluster-score noise of the statistic being assessed. Keeping these objects separate makes clear which conclusions follow from the adoption pattern alone and which require an empirical calibration.

Formally, let $w$ denote the TWFE weights on the $N_1$ treated cells and let $u=1/N_1$ denote the uniform weights defining the equally weighted treated-cell ATT. Under a local drift with heterogeneity scale $c$ and centered, unit-variance profile $s$, the standardized TWFE numerator has mean $\eta=(c/\sigma)(w-u)'s$. Hence
\[
  \max_s|\eta|=\frac{c}{\sigma}\Gamma,
  \qquad \Gamma:=\sqrt{N_1}\|w-u\|_2.
\]
The outcome-free statistic $\Gamma$ records unrestricted design exposure. Projecting $w-u$ onto cohort, event-time, additive cohort-plus-event-time, or saturated group-time profiles gives matched restricted statistics $\Gamma_S$. The saturated equality $\Gamma_{\mathrm{gt}}=\Gamma$ holds when residualized treatment is constant within treated group-time cells, including the balanced unit-and-time-effects designs used below.

For clustered applications, the relevant denominator is the standard deviation of the treatment score, not a separately estimated residual variance and serial-correlation adjustment. We estimate that score scale directly from the reported cluster sandwich and show its consistency as the number of units grows with fixed panel length. The fixed-effect centering identity makes the argument unusually transparent: estimation of the unit effects contributes exactly zero to every treatment score. The result is a structured specialization of the broader many-covariate cluster theory of \citet{AnatolyevNg2026}.

Empirical calibration introduces a second inferential problem. A debiased quadratic estimate of heterogeneity is nonregular when true heterogeneity is zero, so an ordinary percentile interval cannot support a certificate at precisely the boundary where a pass is most plausible. We instead work with the projected vector itself. A multiplier radius and the reverse triangle inequality give a lower bound for detecting an unsafe envelope. For certification, we invert the projected Wald statistic's noncentral-$\chi^2$ law and translate its Mahalanobis upper bound to the Euclidean norm without changing the heterogeneity class. The estimated score denominator enters by ordinary plug-in under the bounded local experiment; no deterministic rate buffer is needed. The two one-sided statements produce three outcomes: certify, withhold certification, or report that the calibration is inconclusive. A signed directional plug-in is reported separately as evidence about whether the estimated effect profile actually points toward the adverse design direction.

The applications are chosen to show why each outcome matters, not to assemble a catalogue of failures. The minimum-wage application in \citet{CS} remains the headline case: every treated-cell weight is positive, yet the conventional interval is not certified. A Brazilian property-tax reanalysis based on \citet{ChristensenGarfias2021} supplies the complementary large-panel result: $17.3\%$ of treated-cell weights are negative and all four classes fail, with a smallest lower bound larger than minimum wage's. Together they show that neither the absence nor the presence of negative weights determines inference adequacy, and that the negative-weight share is not a severity ranking. The castle-doctrine application in \citet{ChengHoekstra} supplies a genuine pass under economically structured heterogeneity and an inconclusive result for the broadest class. The no-fault-divorce application in \citet{StevensonWolfers} shows why a large point envelope need not be decisive when the heterogeneity calibration is noisy and high-dimensional. Section~\ref{sec:apps} develops the substantive context and reports the estimates and confidence bounds.

The paper makes three contributions. First, it converts TWFE weighting geometry into an exact mean-shift bound measured in the units of the reported test, with restricted-profile and directional versions. Second, it shows why positive weights, small coefficient corrections, and sign-reversal diagnostics do not by themselves certify inference: absolute bias can shrink under panel replication while size distortion remains unchanged. Third, it makes the bound feasible by combining a direct fixed-$T$, many-cluster score result with regular, covariance-aware one-sided decisions at zero heterogeneity. The certification step connects the problem to equivalence testing \citep{Romano2005,Lavergne2014} while retaining the Euclidean worst-case norm dictated by the TWFE geometry. Together these results turn a familiar warning about heterogeneous effects into a decision rule that distinguishes what the design certifies from what it merely makes possible. Complete proofs, additional simulations, construction details, and extensions appear in Appendices~\ref{app:proofs}--\ref{app:extensions}.

\subsection{Relation to the literature}

The estimand literature supplies the ingredients but not the size inversion. \citet{dCdH} express TWFE as a weighted average and propose $\bar\sigma_{fe}$, the smallest dispersion of effects capable of reversing the estimand's sign. \citet{GoodmanBacon} locate problematic already-treated comparisons in the two-by-two decomposition, while \citet{Jakiela} turns weight and heterogeneity checks into practitioner diagnostics. These are indispensable estimand objects. Our statistic uses the same weights but standardizes their bias by the sampling noise of the reported test. That change of scale is what allows a design with positive weights and close point estimates to fail.

The remedy literature constructs interpretable group-time and event-time estimators \citep{CS,SunAbraham,BorusyakEtAl}. We use those estimators to calibrate heterogeneity rather than compete with them: a failed certificate directs the researcher to that literature. Parallel-trend diagnostics and sensitivity analyses \citep{Roth,RambachanRoth} concern a different maintained assumption and remain complementary. On inference, unit clustering in DiD follows the warning of \citet{BDM} and the within-estimator result of \citet{Arellano1987}; see \citet{CameronMiller} and \citet{MNW} for practice. The new object is the explicit bias-per-cluster-standard-error threshold, not a new general-purpose variance estimator. This distinction also separates our result from \citet{AnatolyevNg2026}: their contribution is robust variance estimation with many covariates and clustered errors, whereas ours combines a consistent score scale with a heterogeneity-calibrated adequacy decision.

The broader warning that correcting one bias need not repair the combined inferential problem has precedents outside DiD. In nonlinear fixed-effect panels with measurement error, \citet{EvdokimovZeleneev2020Panels} derive cases in which attenuation and incidental-parameter biases have opposite signs: jackknife removal of the $1/T$ component can worsen net bias while leaving the leading measurement-error component intact. In a different measurement-error setting, \citet{MeyerMittag2017} show that imposing an incorrect error--covariate restriction can move corrected estimators farther from a validated target. Our contribution makes the corresponding question operational for TWFE weighting bias by expressing what remains after point-estimate correction in units of the reported test's cluster-score noise.

\section{Theory and diagnostic}\label{sec:setup}

\subsection{Setup and design exposure}

Consider a balanced panel of units $i=1,\ldots,G$ and periods $t=1,\ldots,T$, with staggered absorbing treatment $D_{it}=\mathbf 1\{t\ge g_i\}$. The outcome model is
\begin{equation}\label{eq:model}
 Y_{it}=\alpha_i+\lambda_t+D_{it}\Delta_{it}+\eps_{it}.
\end{equation}
Let $M$ annihilate unit and time indicators, $\Dt=MD$, and $\nw=\Dt'\Dt$. By Frisch--Waugh--Lovell,
\begin{equation}\label{eq:beta}
 \hbeta=\frac{\Dt'Y}{\nw}.
\end{equation}
On the $N_1$ treated cells, define
\begin{equation}\label{eq:weights}
 w_{it}=\frac{\Dt_{it}}{\nw},\qquad u_{it}=\frac1{N_1}.
\end{equation}
Because $\sum_{D=1}\Dt_{it}=\Dt'D=\Dt'\Dt=\nw$, the weights sum to one, although individual weights need not be positive. The TWFE estimand is $\sum_{D=1}w_{it}\Delta_{it}$; the target is $\ATT=N_1^{-1}\sum_{D=1}\Delta_{it}$.

\begin{definition}[Design exposure]\label{def:Gamma}
\[
 \Gamma=\sqrt{N_1}\|w-u\|_2.
\]
For a linear heterogeneity subspace $S\subseteq\R^{N_1}$ containing constants, let $\widetilde S=S\cap\onevec^\perp$ and define
\[
 \Gamma_S=\sqrt{N_1}\|\Pi_{\widetilde S}(w-u)\|_2.
\]
\end{definition}

$\Gamma$ is dimensionless and outcome-free. It vanishes exactly when treated-cell weights are uniform. A negative treated-cell weight is sufficient for $\Gamma>0$, but it is not necessary: the minimum-wage design has no negative weight and $\Gamma>0$. Neither positivity nor the mere inequality $\Gamma>0$ determines whether the adequacy threshold below is crossed.

We use a local drift to obtain a nondegenerate size experiment.

\begin{assumption}[Local heterogeneity]\label{ass:local}
On treated cells,
\[
 \Delta_{it}=\ATT+\frac{c}{\sqrt\nw}s_{it},
 \qquad \onevec's=0,\quad \|s\|_2=\sqrt{N_1},
\]
where $c\ge0$ indexes the magnitude and $s$ the profile of heterogeneity.
\end{assumption}

The normalization makes the cross-cell variance of $s$ equal to one. The $\nw^{-1/2}$ rate is not a claim that economic heterogeneity literally shrinks with a larger sample; it is the local experiment in which weighting bias remains nondegenerate relative to the standard error, just as local-to-zero asymptotics isolate weak-instrument distortions \citep{SS,SY}.

\begin{theorem}[Exact numerator and worst-case mean shift]\label{thm:iid}
Suppose $\eps\mid D\sim N(0,\sigma^2I)$ and Assumption~\ref{ass:local} holds. Conditional on the design,
\[
 \frac{\sqrt\nw(\hbeta-\ATT)}{\sigma}\sim N(\eta,1),
 \qquad \eta=\frac c\sigma(w-u)'s.
\]
For any nontrivial $\widetilde S$,
\begin{equation}\label{eq:worst}
 \max_{s\in\widetilde S,\ \|s\|=\sqrt{N_1}}|\eta|
 =\frac c\sigma\Gamma_S,
\end{equation}
If $\Pi_{\widetilde S}(w-u)\ne0$, a maximizer is proportional to that projection. If the projection is zero, the maximum is zero and every admissible profile is a maximizer. The unrestricted case gives $(c/\sigma)\Gamma$.
\end{theorem}

The conventional residual variance adds no first-order distortion. More precisely, with $Q=M-\Dt\Dt'/\nw$, the residual heterogeneity is $h=(c/\sqrt\nw)Q(D\circ s)$. The numerator and residual variance are independent under Gaussianity, and the latter is a noncentral chi-square with noncentrality $\|h\|^2/\sigma^2$. It is therefore stochastically no smaller than its central counterpart: the corresponding noncentral-$t$ rejection probability is a finite-sample upper bound, and the studentization difference vanishes under the usual design regularity. Appendix~\ref{app:proofs} gives the exact statement and proof.

The relevant subspaces have an immediate interpretation. $S_{\mathrm{coh}}$ consists of profiles constant within adoption cohort; $S_{\mathrm{evt}}$ consists of profiles constant at each event time; $S_{\mathrm{c+e}}=S_{\mathrm{coh}}+S_{\mathrm{evt}}$ contains additive cohort and dynamic effects; and $S_{\mathrm{gt}}$ is saturated in post-treatment cohort-by-calendar-time cells. Orthogonal-projection monotonicity gives
\begin{equation}\label{eq:nesting}
 \Gamma_{\mathrm{coh}},\Gamma_{\mathrm{evt}}
 \le \Gamma_{\mathrm{c+e}}\le\Gamma_{\mathrm{gt}}\le\Gamma.
\end{equation}
In a balanced staggered panel with unit and time effects and no additional observation-varying controls, $\Dt_{it}$ depends on a treated observation only through its cohort and calendar time. More generally, the following equality requires residualized treatment to be constant within every treated group-time cell. Under that condition, $w-u\in S_{\mathrm{gt}}\cap\onevec^\perp$ and
\begin{equation}\label{eq:gtclosure}
 \Gamma_{\mathrm{gt}}=\Gamma.
\end{equation}
Thus the saturated group-time class contains the entire weighting-relevant adverse direction; within-cell effect variation is orthogonal to $w-u$. Each bound is calibrated by a pilot for the same component of heterogeneity.

\subsection{Why estimand diagnostics do not certify inference}\label{sec:scale}

Replicate every unit in the design $K$ times, preserving its adoption path and local profile. Then $N_1$ and $\nw$ both multiply by $K$, each weight deviation divides by $K$, and $\Gamma_S$ is unchanged. At fixed $c/\sigma$, the mean shift and limiting rejection probability are therefore unchanged. But
\begin{equation}\label{eq:bias}
 |\E\hbeta-\ATT|=\frac{|\eta|\sigma}{\sqrt\nw}=O(K^{-1/2}).
\end{equation}
An absolute-bias diagnostic becomes more reassuring while the test remains equally distorted. Under a different thought experiment that holds per-cell economic dispersion fixed, $c$ grows like $\sqrt\nw$ and both diagnostics eventually flag. The nonredundancy result concerns the local experiment that defines a stable size threshold, not every possible replication scheme.

This distinction can also be expressed in the units of the estimand diagnostics. Let $\sigma_\Delta^2=N_1^{-1}\sum_{D=1}(\Delta_{it}-\ATT)^2$. Cauchy--Schwarz gives
\begin{equation}\label{eq:signreverse}
 |\beta_{TWFE}-\ATT|
 \le \|w-u\|\sqrt{N_1}\sigma_\Delta
 =\Gamma\sigma_\Delta.
\end{equation}
Consequently $\bar\sigma_0=|\beta_{TWFE}|/\Gamma$ is the minimum treated-cell RMS dispersion capable of moving an equally weighted effect to zero, the direct sign-reversal calculation underlying the dispersion diagnostic of \citet{dCdH}. It asks whether heterogeneity could overturn the sign of the target. The certificate below instead asks whether admissible weighting bias could be large relative to the reported standard error. Neither answer determines the other.

\subsection{Direct feasible cluster calibration}\label{sec:cluster}

Let $d_i$ be unit $i$'s $T$-vector of within-treatment values and allow arbitrary within-unit dependence, with clusters independent and $\E(\eps_i\mid D)=0$. The two-way transformation gives the exact identity
\begin{equation}\label{eq:center}
 d_i'\onevec=0\qquad\text{for every }i.
\end{equation}
Let $S_i=d_i'\eps_i$ and define the cluster-score noise scale
\begin{equation}\label{eq:q}
 q_G^2=\frac1\nw\sum_{i=1}^G\Var(S_i\mid D),
 \qquad
 \Var(\hbeta\mid D)=\frac{q_G^2}{\nw}.
\end{equation}
If $\Cov(\eps_i\mid D)=\sigma^2R_i$, then $q_G^2=\sigma^2\psi_G$ with $\psi_G=\nw^{-1}\sum_i d_i'R_i d_i$. Thus the earlier oracle notation $\Gamma_{S,CR}=\Gamma_S/\sqrt\psi$ is simply an interpretive factorization of $c\Gamma_S/q$.

To pin down the finite-sample convention, define the cluster-only corrected score sandwich
\begin{equation}\label{eq:cr1}
 \widehat V_C=\frac{G}{G-1}\frac1{\nw^2}
 \sum_{i=1}^G(d_i'\widehat\eps_i)^2,
 \qquad
 \qhat^2:=\nw\widehat V_C.
\end{equation}
Throughout, ``direct score scale'' means exactly~\eqref{eq:cr1}. It includes $G/(G-1)$ but not an additional $(n-1)/(n-K)$ factor; the latter does not approach one when unit effects proliferate with fixed $T$ and would define a different finite-sample convention.

\begin{assumption}[Many clusters, fixed panel length]\label{ass:cluster}
$T$ is fixed and $G\to\infty$; clusters are independent with $\E(\eps_i\mid D)=0$; $\sup_i\E\|\eps_i\|^{4+\delta}<\infty$ for some $\delta>0$; $\nw/G$ and $q_G^2$ are bounded above and away from zero; $\max_i\|d_i\|^2/\nw\to0$; and the cluster scores satisfy the corresponding Lyapunov condition. The number of common time effects is fixed, and the local scale $c$ is bounded.
\end{assumption}

\begin{theorem}[Fixed-$T$ feasible cluster statistic]\label{thm:feasible}
Under Assumptions~\ref{ass:local} and~\ref{ass:cluster},
\[
 \qhat^2-q_G^2=O_p(G^{-1/2}),
 \qquad
 \frac{\hbeta-\ATT}{\widehat{\mathrm{SE}}_C}
 \ \Rightarrow\ N(\eta_{CR},1),
 \qquad
 \eta_{CR}=\lim_G\frac c{q_G}(w-u)'s,
\]
whenever the displayed limit exists. Consequently the worst-case limiting mean shift over $S$ is $c\Gamma_S/q_G$.
\end{theorem}

The result does not require $T\to\infty$. Equation~\eqref{eq:center} makes the estimated unit effect disappear from each score exactly. Estimation of the fixed number of time effects perturbs the normalized score meat by $O_p(G^{-1/2})$; slope estimation and the bounded local residual heterogeneity are of the same or smaller order. The fourth-moment bound gives the root-$G$ rate for the average of $S_i^2$. Appendix~\ref{app:proofs} provides the decomposition. \citet{AnatolyevNg2026} establish a broader cluster-and-many-covariate theory, including growing and unbalanced clusters and variance estimators robust to many covariates. Theorem~\ref{thm:feasible} exploits the fixed-effect centering identity to deliver the narrower fixed-$T$ bridge used here. Any alternative standard error with the same relative consistency can replace $\widehat{\mathrm{SE}}_C$ in the one-sided procedures below. This variance-estimator statement does not extend the Gaussian rejection map mechanically to finite-$G$ $t$ or bootstrap tests.

\subsection{Threshold, empirical pilot, and audit}\label{sec:threshold}

For a two-sided nominal-$\alpha$ Gaussian test, define
\begin{equation}\label{eq:size}
 p(\eta)=\Phi(-z_{1-\alpha/2}-\eta)+1-\Phi(z_{1-\alpha/2}-\eta).
\end{equation}
It is even and strictly increasing in $|\eta|$, with
\[
 p(\eta)=\alpha+z_{1-\alpha/2}\phi(z_{1-\alpha/2})\eta^2+O(\eta^4).
\]
For an allowed size distortion $\delta$, let $\eta^\dagger(\alpha,\delta)$ solve $p(\eta)=\alpha+\delta$. At $\alpha=\delta=0.05$, $\eta^\dagger=0.652358$ (the quadratic approximation is $0.661$). Hence the population adequacy condition in subspace $S$ is
\begin{equation}\label{eq:threshold}
 \frac{c_S\Gamma_S}{q_G}\le\eta^\dagger.
\end{equation}
\companionline

To calibrate $c_S$, stack heterogeneity-robust group-time effects $\widehat\Delta$ and let $A$ map them to the $N_1$ treated cells. The pilot requires more than a regression calculation: no anticipation and the stated comparison-group parallel-trends restriction must identify every included effect, each treated cohort and comparison pool must have nonvanishing asymptotic share, and $A$ must cover every target treated cell. We take the finite group-time dimension and the subspace $S$ as fixed before inspecting outcomes. Write $P_S=\Pi_{\widetilde S}$ and
\begin{equation}\label{eq:mhat}
 \widehat m_S=\sqrt{\frac{\nw}{N_1}}P_SA\widehat\Delta,
 \qquad m_S=\sqrt{\frac{\nw}{N_1}}P_SA\Delta,
 \qquad c_S=\|m_S\|.
\end{equation}
Because $P_S\onevec=0$, explicit recentering is unnecessary. With $\Omega=\Var(\widehat\Delta)$, a covariance-debiased point calibration is
\begin{equation}\label{eq:pilot}
 (\chat^{\mathrm{db}}_S)^2=\max\left\{0,
 \|\widehat m_S\|^2-
 \frac{\nw}{N_1}\tr(P_SA\widehat\Omega A'P_S)
 \right\}.
\end{equation}
The trace removes the expected projected estimation noise. Replacing it by an average marginal variance is generally wrong because group-time estimates share comparison observations. We use
\begin{equation}\label{eq:pointenvelope}
 \Khat^{\mathrm{pt}}_S=\frac{\chat^{\mathrm{db}}_S\Gamma_S}{\qhat}
\end{equation}
as a point estimate of the worst-case mean shift and label $p(\Khat^{\mathrm{pt}}_S)$ a point envelope. The truncation in~\eqref{eq:pilot} is nonregular at $c_S=0$, so neither a percentile bootstrap of $\chat^{\mathrm{db}}_S$ nor its point value supplies a confidence certificate.

Inference instead operates on the vector before taking its norm. In the applications $\widehat\Delta$ is the explicit long-difference ATT$(g,t)$ estimator, and the exact cluster influence scores estimate both its covariance and its multiplier law; the saturated mean surface is used only to estimate residuals. Let $\widehat\Delta^*$ be a cluster multiplier draw, centered at its fitted conditional mean, and let $\widehat r_{1-\gamma,S}$ be the conditional $(1-\gamma)$ quantile of
\begin{equation}\label{eq:radius}
 R_S^*=\sqrt{\frac{\nw}{N_1}}
 \left\|P_SA(\widehat\Delta^*-\widehat\Delta)\right\|.
\end{equation}
The implementation uses HC2-scaled residual scores and reports HC3 as a finite-sample sensitivity. These adjustments apply to inference on the heterogeneity pilot, not to the conventional TWFE test being diagnosed. The reverse triangle inequality gives the one-sided lower bound
\begin{equation}\label{eq:lower}
 \boxed{\quad
 \Khat^-_S=\frac{\Gamma_S}{\qhat}
   (\|\widehat m_S\|-\widehat r_{1-\gamma,S})_+.
 \quad}
\end{equation}

Certification uses the covariance rather than adding a full-vector radius. Let $Q_S$ have orthonormal columns spanning the image of $P_S$, set $\widehat x_S=(\Gamma_S/\qhat)Q_S'\widehat m_S$, and let $\widehat C_S$ be its score covariance. With $d_S=\dim(\widetilde S)$, define
\[
 W_S=\widehat x_S'\widehat C_S^{-1}\widehat x_S.
\]
For $W\sim\chi^2_{d_S}(\lambda)$, let $\widehat\lambda^+_S$ solve $\Pr\{W\le W_S\}=\gamma$; set it to zero if the central cdf at $W_S$ is already at most $\gamma$. The covariance-aware Euclidean upper bound is
\begin{equation}\label{eq:upper}
 \boxed{\quad
 \Khat^+_S=
 \sqrt{\lambda_{\max}(\widehat C_S)\widehat\lambda^+_S}.
 \quad}
\end{equation}
The inequality $\|x_S\|^2\le\lambda_{\max}(C_S)x_S'C_S^{-1}x_S$ preserves the original Euclidean worst-case target. It can be conservative when the covariance is strongly anisotropic, but it avoids requiring an entire confidence ball to fit inside the adequacy region. If $\widehat C_S$ is rank deficient relative to the prespecified subspace, the implementation issues no certificate rather than regularizing an unidentified direction.

\begin{theorem}[Regular one-sided decisions]\label{thm:certificate}
Suppose the cluster multiplier consistently estimates the distribution of $\|\widehat m_S-m_S\|$, its $(1-\gamma)$ quantile is continuous, $c_S$ is bounded along the local sequence, $q_G$ is bounded away from zero, and $\qhat/q_G\to_p1$. Suppose also that the fixed-dimensional covariance estimator $\widehat C_S\to_p C_S$, where $C_S$ is positive definite on the image of $P_S$. Then, writing $K_S=c_S\Gamma_S/q_G$,
\[
 \liminf_G\Pr\{\Khat^-_S\le K_S\}\ge1-\gamma,
 \qquad
 \liminf_G\Pr\{K_S\le\Khat^+_S\}\ge1-\gamma.
\]
Consequently, certifying when $\Khat^+_S\le\eta^\dagger$ has asymptotic false-certification probability at most $\gamma$ whenever $K_S>\eta^\dagger$. Withholding certification when $\Khat^-_S>\eta^\dagger$ has the analogous error control when $K_S<\eta^\dagger$. These are separate one-sided $1-\gamma$ guarantees, not a joint $1-\gamma$ confidence interval, and both include $c_S=0$.
\end{theorem}

The lower-bound proof uses
\[
 \big|\|\widehat m_S\|-c_S\qhat/q_G\big|
 \le \|\widehat m_S-m_S\|+c_S|\qhat/q_G-1|.
\]
The second term is $o_p(1)$ under bounded local heterogeneity, so quantile continuity absorbs denominator estimation without a deterministic rate buffer. For the upper bound, the Wald statistic has a noncentral-$\chi^2$ limit, noncentrality $x_S'C_S^{-1}x_S$, and noncentrality inversion is one-sided valid; the eigenvalue inequality completes the translation. This is the multivariate equivalence-testing logic of \citet{Lavergne2014}, adapted to the Euclidean target fixed by the TWFE weighting geometry; see also \citet{Romano2005}. Algebraically, the point envelope still equals $(\chat_S^{\mathrm{db}}/\widehat\sigma)\Gamma_S/\sqrt{\widehat\psi}$ when $\widehat\psi=\qhat^2/\widehat\sigma^2$. Because $\widehat\sigma$ cancels, the direct form avoids separately interpreting a residual variance and a working serial-correlation model.

A more general confidence-set construction remains available. If $a_G\downarrow0$ satisfies $\sqrt G a_G\to\infty$, combine the multiplier norm radius with the denominator band $|\qhat/q_G-1|\le a_G$. This produces a joint interval by multiplying the reverse-triangle lower endpoint by $1-a_G$ and the full-ball upper endpoint by $1+a_G$. It does not use boundedness of $c_S$ in the denominator step and does not require a nonsingular Gaussian covariance for an upper statement. Its cost is power: the full confidence ball can be too wide to fit inside the adequacy region even when heterogeneity is zero. It is the relevant fallback when $q_G$ is estimated by an unusual scheme that calls for an explicit rate band, or when a nonsingular Gaussian projected covariance cannot be maintained. We use the decision-specific rule in Theorem~\ref{thm:certificate} because the maintained application experiment is bounded and fixed-dimensional and the empirical question is equivalence to a stated margin. The Online Supplement states the alternative formally and explains the tradeoff. Appendix~\ref{app:normbound} supplies primitive fixed-$T$ conditions and the exact influence-score construction.

The population inequality is deliberately one-sided in its interpretation. If $c_S\Gamma_S/q_G\le\eta^\dagger$, the calibrated magnitude is safe for every profile in $S$. If the envelope exceeds $\eta^\dagger$, the class contains a profile that violates the tolerance. That fact does not identify the shift generated by the true profile. To measure orientation, define
\[
 h_S=P_S(A\Delta-\overline{A\Delta}\onevec),\qquad
 r_S=\frac{\langle P_S(w-u),h_S\rangle}
 {\|P_S(w-u)\|\,\|h_S\|}.
\]
Whenever both norms are positive, $r_S\in[-1,1]$ and the signed shift generated by this profile is
\begin{equation}\label{eq:directional}
 \eta_{\mathrm{dir},S}
 =\frac{\sqrt\nw}{q}(w-u)'h_S
 =r_S\frac{c_S\Gamma_S}{q}.
\end{equation}
Thus the worst case is an envelope and $r_S$ records the fraction used by a given profile, including its sign.

The empirical directional companion is the linear contrast
\begin{equation}\label{eq:directionalhat}
 \widehat\eta_{\mathrm{dir},S}
 =\frac{\sqrt\nw}{\qhat}(w-u)'
 P_S(A\widehat\Delta-\overline{A\widehat\Delta}\onevec)
 =\frac{(w-u)'P_S(A\widehat\Delta-\overline{A\widehat\Delta}\onevec)}
 {\widehat{\mathrm{SE}}_C}.
\end{equation}
Unlike the squared-norm pilot, this signed linear contrast needs no trace subtraction when $\widehat\Delta$ is unbiased. It does need a bootstrap interval because estimated orientation is noisy. Since $\Khat_S$ uses a noise-corrected norm while~\eqref{eq:directionalhat} is linear, their sample analogues need not satisfy $\widehat\eta_{\mathrm{dir},S}=\widehat r_S\Khat_S$ exactly.

\begin{framed}
\begin{spacing}{1.15}
\noindent\textbf{Five-step audit.}
\begin{enumerate}[label=\arabic*.,nosep]
\item Residualize treatment on the exact fixed effects and controls; compute $w$, $\Gamma$, and the matched $\Gamma_S$ ladder, including $S_{\mathrm{gt}}$.
\item Estimate heterogeneity-robust group-time effects, map them to treated cells, and compute $\widehat m_S$ and the debiased point calibration~\eqref{eq:pilot}.
\item Compute the exact score sandwich~\eqref{eq:cr1} and set $\qhat=\sqrt\nw\widehat{\mathrm{SE}}_C$.
\item Use the exact ATT$(g,t)$ cluster scores to obtain the lower bound~\eqref{eq:lower} and the covariance-aware upper bound~\eqref{eq:upper}; compare the separate one-sided bounds with $\eta^\dagger$ and report any covariance-rank failure.
\item Label $p(\Khat_S^{\mathrm{pt}})$ a point worst-case envelope and report the directional plug-in~\eqref{eq:directionalhat} separately. A failed certificate is not a claim that the true effect profile attains the envelope.
\end{enumerate}
\end{spacing}
\end{framed}

\section{Applications and empirical verdicts}\label{sec:apps}

The empirical analysis revisits four studies that became reference points in applied economics or the development of modern panel methods. They concern the employment effects of minimum wages, municipal fiscal capacity in Brazil, the consequences of expanded self-defense laws, and the effects of divorce reform on family distress. These are not interchangeable demonstration datasets. Each asks a consequential policy question, and each supplies a different combination of weighting exposure, outcome heterogeneity, cluster support, and covariance rank. Together they test whether the diagnostic can distinguish different evidentiary situations rather than simply find reasons for concern.

The exercise is deliberately narrower than a complete reanalysis of any of the four literatures. We take the public panels, treatment timing, outcomes, and conventional TWFE specifications as given, and use the equally weighted treated-cell ATT as the common target. We do not reassess every identifying assumption or claim that the conventional regression is the original authors' preferred estimator. The question is instead counterfactual and diagnostic: if a researcher reported the conventional clustered TWFE test, what can the design and the observed heterogeneity certify about its size?

For the broadest calculation, we use the saturated post-treatment group-time class. In these balanced specifications $\Gamma_{\mathrm{gt}}=\Gamma$, so this class contains the full weighting-relevant adverse direction. Cohort, event-time, and additive cohort-plus-event-time results show how the answer changes under lower-dimensional economic structure; the signed directional calculation uses the additive fit. Table~\ref{tab:headline} separates point envelopes from the one-sided bounds that determine the verdict. Minimum wage cannot be certified even in the additive class. Brazil withholds certification in every class despite strong sample support. Castle-doctrine homicide is certified in the additive class but inconclusive when saturated, and divorce is inconclusive.

\begin{table}[htbp]
\centering
\begin{threeparttable}
\caption{Application diagnostics}\label{tab:headline}
\small
\begin{tabular}{lrrrrr}
\toprule
\multicolumn{6}{l}{\textit{Panel A: estimand and design diagnostics}}\\
& clusters $\times T$ & TWFE & neg. wt. & $\Gamma$ & $\bar\sigma_0$ \\
\midrule
Minimum wage & $2{,}284\times7$ & $-0.0366$ & $0.0\%$ & $0.270$ & $0.136$ \\
Brazil property tax & $2{,}840\times12$ & $0.0593$ & $17.3\%$ & $1.138$ & $0.0521$ \\
Castle doctrine & $50\times11$ & $0.0818$ & $0.0\%$ & $0.211$ & $0.387$ \\
No-fault divorce & $49\times27$ & $-0.0158$ & $1.2\%$ & $0.643$ & $0.0245$ \\
\addlinespace
\multicolumn{6}{l}{\textit{Panel B: inference diagnostics by reported class}}\\
& class & $\Gamma_S$ & $\Khat^{\mathrm{pt}}_S$ & one-sided $95\%$ result & decision \\
\midrule
Minimum wage & group-time & $0.270$ & $1.480$ & $K^-\!=0.793$ & withhold \\
Brazil property tax & cohort & $0.463$ & $3.053$ & $K^-\!=1.040$ & withhold \\
Castle doctrine & additive & $0.198$ & $0.000$ & $K^+\!=0.236$ & certify \\
Castle doctrine & group-time & $0.211$ & $0.086$ & covariance rank $18/19$ & inconclusive \\
No-fault divorce & group-time & $0.643$ & $2.485$ & covariance rank $49/167$ & inconclusive \\
\bottomrule
\end{tabular}
\begin{tablenotes}[flushleft]\footnotesize
\item Notes: $\bar\sigma_0=|\widehat\beta_{TWFE}|/\Gamma$ is the minimum treated-cell RMS effect dispersion capable of moving the equally weighted effect to zero. $\Khat^{\mathrm{pt}}_S$ uses the trace-debiased calibration~\eqref{eq:pilot}; it is a worst-case point envelope, not a realized-size estimate. Brazil reports the cohort class because it has the smallest lower bound; all four classes withhold certification. Lower bounds use the HC2 multiplier radius and reverse triangle inequality. Upper bounds invert the noncentral-$\chi^2$ law of the projected Wald statistic and translate its Mahalanobis bound to the Euclidean target. Each is a separate one-sided $95\%$ statement. No upper certificate is issued when the score covariance is rank deficient relative to the prespecified subspace. The adequacy threshold is $0.652$, corresponding to a $10\%$ size ceiling.
\end{tablenotes}
\end{threeparttable}
\end{table}

\subsection{Minimum wages: positive weights do not certify inference}\label{sec:minwage}

Few empirical questions have been as durable in labor economics as whether minimum wages reduce employment. The natural-experiment evidence of \citet{CardKrueger1994} challenged the then-prevailing empirical consensus and helped place research design at the center of the debate. Later border-based work made the choice of geographic comparison groups equally prominent \citep{DubeLesterReich2010}. Minimum wages have consequently become a proving ground not only for substantive claims about labor demand, but also for new methods of policy evaluation.

The application in \citet{CS} matters in both respects. It revisits teen employment using county-level policy variation, and it serves as the principal empirical illustration of a framework that separates group-time effects from the researcher's chosen aggregation. Our audit is not a criticism of their recommended procedure: they already use heterogeneity-robust estimation and clustered bootstrap inference. That makes the setting especially informative. It lets us ask whether two seemingly reassuring facts about the conventional regression (positive weights and a moderate point-estimate correction) would independently justify confidence in its usual interval.

Their application studies county teen employment from 2001 through 2007. Counties belong to states that first raised their minimum wage in 2004, 2006, or 2007, or to states whose effective minimum remained at the federal level. The exact balanced replication sample contains 100, 223, and 584 counties in those adoption groups and 1,377 never-treated counties, generating 1,430 treated county-year cells. Their main unconditional analysis uses never-treated counties as controls, no anticipation, and county clustering. We use the same treatment timing, outcome, and $2,284$ counties.

The unconditional TWFE regression reproduces the published coefficient: $-0.0366$ here versus $-0.037$ in their Table~3. Their reported $-0.039$ group aggregate and the target in this paper are not the same estimand: the former aggregates cohorts, while the latter weights treated county-year cells equally. Weighting the seven heterogeneity-robust group-time effects for the latter target gives $-0.0516$. That is the relevant comparison, and we use it throughout. The separate weight audit remains superficially reassuring (as no treated-cell weight is negative) but the positive weights are not uniform. The design statistics are
\[
 (\Gamma_{\mathrm{coh}},\Gamma_{\mathrm{evt}},
 \Gamma_{\mathrm{c+e}},\Gamma_{\mathrm{gt}},\Gamma)
 =(0.166,0.251,0.257,0.270,0.270).
\]
Nearly all of the imbalance is already aligned with additive cohort and dynamic heterogeneity, and the saturated group-time projection captures it exactly.

These calculations locate the estimand issue but do not yet answer the inference question. Equation~\eqref{eq:signreverse} gives $\bar\sigma_0=0.1358$: treated-cell effects would need at least $13.6$ log points of RMS dispersion before weighting alone could move the equally weighted effect to zero. The $1.50$-log-point gap between the TWFE and target-matched robust estimates is one weighted contrast and does not estimate or upper-bound that RMS dispersion. The sign-reversal threshold and the observed coefficient gap therefore answer different questions; their numerical comparison cannot establish that reversal is remote.

The inferential audit begins where that estimand comparison ends. The seven post-treatment group-time effects supply the heterogeneity pilot. We subtract their full projected sampling covariance, rather than an average of marginal standard errors, because the estimates share the same never-treated comparison units. The trace-debiased saturated scale is $\chat^{\mathrm{db}}_{\mathrm{gt}}/\widehat\sigma=6.460$. The direct score ratio is $\widehat\psi=\qhat^2/\widehat\sigma^2=1.385$, giving
\[
 \Khat^{\mathrm{pt}}_{\mathrm{gt}}
 =\frac{6.460\times0.270}{\sqrt{1.385}}
 =1.480.
\]
This point worst-case shift maps to a $31.6\%$ envelope for the nominal-$5\%$ test. The inferential verdict does not use a percentile interval for this truncated quadratic. In $999$ fixed-design county multiplier draws, the HC2 reverse-triangle lower bound is $\Khat^-_{\mathrm{gt}}=0.793$, which maps to a $12.5\%$ size envelope and exceeds the $0.652$ threshold. The additive lower bound is $0.756$ and also exceeds the threshold. HC3 gives a saturated lower bound of $0.792$. The conclusion therefore does not depend on allowing unrestricted group-time heterogeneity: conventional inference is withheld under both the additive and saturated classes.

The estimated profile is not oriented arbitrarily. Its cosine with the adverse additive design direction is $\widehat r=0.925$. The signed plug-in in~\eqref{eq:directionalhat} is $\widehat\eta_{\mathrm{dir}}=1.370$, which maps to $27.8\%$ rejection; the direct wild-bootstrap median is $23.5\%$ and the $95\%$ interval is $[11.8,40.5]\%$. This close agreement makes the envelope economically relevant rather than a visibly remote construction. It remains a plug-in calculation based on estimated group-time effects, not a direct observation of the true rejection probability.

The conclusion is therefore precise. It is not a claim that the sign or substantive conclusion of the original study is wrong, nor that its published robust inference is invalid. \citet{CS} already use heterogeneity-robust estimation and clustered bootstrap inference. The new fact is that positive weights and a comfortable sign-reversal threshold do not certify the conventional TWFE interval. A moderate coefficient correction can still be large in standard-error units in a panel with more than two thousand clusters.

\subsection{Brazilian property taxation: a sharp failure}\label{sec:brazil}

Fiscal cadastres record property ownership and values and can expand local governments' capacity to collect property taxes. \citet{ChristensenGarfias2021} study Brazilian municipalities and report that property-tax revenue rises by more than $10\%$ after a cadastre update. We examine the unit-and-year-effects specification for log property-tax revenue underlying their Table~2, column~1. The source file comes from the public replication archive assembled by \citet{ChiuEtAl2026}. To give every group-time effect an untreated baseline and a complete time path, we use its full-date balanced 2004--2015 subset and exclude municipalities already treated in 2004. The resulting sample has $2{,}840$ municipalities. This is a transparent reanalysis of the published specification on a balanced subset, not an exact reproduction of the full published estimation sample.

The conventional coefficient on the balanced subset is $0.05935$, with a standard error of $0.02679$ clustered by municipality; the equally treated-cell-weighted long-difference aggregate is $0.02846$. Unlike the minimum-wage design, $17.34\%$ of treated-cell weights are negative. Its design ladder is
\[
 (\Gamma_{\mathrm{coh}},\Gamma_{\mathrm{evt}},
  \Gamma_{\mathrm{c+e}},\Gamma_{\mathrm{gt}},\Gamma)
 =(0.463,1.025,1.065,1.138,1.138).
\]
The application therefore complements the title example rather than replacing it: minimum wage establishes that positive weights do not suffice, while Brazil asks whether a failure remains visible in a large, well-supported design where nonuniform and negative weights are more apparent.

It does. The HC2 reverse-triangle lower bounds are $1.040$ for cohort heterogeneity, $2.441$ for event-time heterogeneity, $3.342$ for the additive class, and $3.223$ for the saturated group-time class. Every value exceeds the $0.652$ adequacy threshold; HC3 changes them only to $1.036$, $2.434$, $3.334$, and $3.210$. The cohort bound is the least adverse result and still clears the threshold by $0.388$. This is a substantially sharper noncertification than in minimum wage, though it does not carry the paper's positive-weight claim.

The sample diagnostics make the numerical sharpness interpretable. The smallest treated cohort contains $88$ municipalities, every included comparison pool contains at least $99$, maximum group-time leverage is $0.0101$, and the projected HC2 and HC3 covariance matrices have full rank in all four classes. These facts remove the thin-support and rank concerns that limit the state applications. They do not validate parallel trends or turn a failed uniform certificate into a claim about the realized rejection probability. They show that this particular failure is not a consequence of an imprecise heterogeneity calibration or an unsupported covariance direction.

\subsection{Castle doctrine: a structured certificate}\label{sec:castle}

The third application turns to a stark question in the economics of crime: does expanding the legal protection for lethal self-defense deter offenders, or does it escalate violence? \citet{ChengHoekstra} found that castle-doctrine expansions did not deter burglary, robbery, or aggravated assault and were associated with an approximately eight-percent increase in homicide. The result is substantively important because the two competing mechanisms imply opposite policy conclusions. It also provides an essential counterweight for the diagnostic. A useful certification procedure must sometimes find the conventional analysis adequate, rather than merely attach a warning to every staggered design.

The public state-year panel contains $50$ states from 2000 through 2010, with 21 adopters and 29 never-treated states. The TWFE estimate is $0.0818$, close to the study's reported homicide increase. Again no treated-cell weight is negative. The design is closer to uniform than minimum wage,
\[
 (\Gamma_{\mathrm{coh}},\Gamma_{\mathrm{evt}},
 \Gamma_{\mathrm{c+e}},\Gamma_{\mathrm{gt}},\Gamma)
 =(0.142,0.168,0.198,0.211,0.211),
\]
but the decisive difference is outcome-side. After subtracting the full covariance of the group-time estimates, the cohort, event-time, and additive pilots hit the nonnegative noise floor. The saturated pilot does not: $\chat^{\mathrm{db}}_{\mathrm{gt}}/\widehat\sigma=0.746$, but its direct point shift is only $0.086$, corresponding to a $5.1\%$ point envelope.

The estimand-side diagnostics also point away from a problem: $\bar\sigma_0=0.387$, and no treated-cell weight is negative. More importantly, the boundary calls for a valid one-sided equivalence test rather than a literal claim of zero heterogeneity. Noncentral-$\chi^2$ inversion gives the additive class an HC2 upper bound of $\Khat^+_{\mathrm{c+e}}=0.236$, comfortably below the $0.652$ threshold; HC3 gives $0.225$. The saturated score covariance has rank $18$ while its centered subspace has rank $19$. We therefore issue no saturated upper certificate instead of silently regularizing the unsupported direction.

The split is useful guidance. With few clusters and thin adoption cohorts, a structured heterogeneity class can be estimable well enough to certify while a saturated surface is not. Here the additive restriction is economically interpretable and clears the threshold with margin, whereas the observed score covariance cannot support inference in every saturated direction. The smallest adoption cohort contains one state and maximum leverage of the residualized treated group-time surface is $0.891$. The additive pass remains conditional on the maintained many-cluster approximation; the HC2 and HC3 calculations are sensitivities, not finite-$G$ coverage theorems.

Here the asymmetry of the method is useful. An upper bound below the ceiling certifies every orientation in the prespecified additive class with the stated asymptotic confidence. The directional additive calculation is consistent with that result: $\widehat\eta_{\mathrm{dir}}=-0.013$, and its descriptive wild-bootstrap size interval is $[5.0,5.5]\%$. The saturated result also shows why the class, denominator uncertainty, and finite-sample multiplier convention must be reported rather than hidden behind a binary label.

Homicide-additive remains the headline certificate because homicide is the central adverse outcome in the study and additive cohort-plus-event-time heterogeneity is the broadest structured class that passes. A family-adjusted outcome audit in the Online Supplement supplies corroboration without changing that headline. Burglary certifies in the prespecified cohort and event-time classes, robbery certifies in those same two classes, and assault certifies in the event-time class. These statements survive a Bonferroni adjustment over the $20$ recorded outcomes crossed with four classes; assault-cohort does not. The pattern makes the restriction transparent: the additional passes are evidence for narrower economic structures, not saturated certificates.

\subsection{No-fault divorce: an inconclusive point flag}\label{sec:divorce}

Unilateral-divorce reforms changed the legal outside option within marriage and became an important setting for studying household bargaining. \citet{StevensonWolfers} used the reforms to ask whether easier access to divorce reduced family distress, including female suicide and domestic violence. The same panel later became a central illustration in \citet{GoodmanBacon}'s account of the comparisons embedded in staggered-adoption TWFE. Its substantive importance and methodological afterlife make it a natural stress test for the distinction between an alarming point calculation and an evidentially supported verdict.

The state-year panel covers 1964--1990. Two states already treated in 1964 have no in-sample untreated baseline and are excluded, leaving $49$ states, $20$ never-treated states, and $522$ treated cells. Including those states under an extended convention raises $\Gamma$, so the exclusion is conservative for the design flag. The design ladder is
\[
 (\Gamma_{\mathrm{coh}},\Gamma_{\mathrm{evt}},
 \Gamma_{\mathrm{c+e}},\Gamma_{\mathrm{gt}},\Gamma)
 =(0.38096,0.46809,0.56236,0.64326,0.64326),
\]
and $1.15\%$ of treated-cell weights are negative.

The estimand diagnostic is already uneasy: $\bar\sigma_0=0.0245$, so comparatively little treated-cell effect dispersion can move the equally weighted effect to zero. This agreement between the estimand and inference diagnostics is informative, but it does not remove uncertainty from either calculation.

The trace-debiased saturated pilot is $\chat^{\mathrm{db}}_{\mathrm{gt}}/\widehat\sigma=6.5994$. With the direct residual-score ratio $\widehat\psi=2.9172$,
\[
 \Khat^{\mathrm{pt}}_{\mathrm{gt}}
 =\frac{6.5994\times0.64326}{\sqrt{2.9172}}
 =2.4855,
\]
which gives a $70.0\%$ point worst-case envelope. The additive point shift remains $1.7447$, corresponding to $41.5\%$. The estimated additive profile has alignment $\widehat r=0.568$ and directional shift $1.629$, mapping to $37.1\%$ rejection.

Sampling uncertainty changes the conclusion. The saturated reverse-triangle lower bound is only $0.083$, below the $0.652$ threshold, while its score covariance has rank $49$ against the $167$ directions required for a saturated upper certificate. The additive upper bound is also far above the threshold. The direct directional size interval is broad, from $5.0\%$ to $47.6\%$. Neither the uniform nor directional calculation establishes rejection above the adequacy ceiling, and the rank diagnostic rules out an unsupported saturated pass.

This application is also the least comfortable match to the asymptotic assumptions. It has only $G=49$ independent state clusters and $T=27$ observations per cluster. The pilot contains 168 identified group-time effects, so the centered saturated projection has rank 167; several adoption cohorts contain a single state, and the maximum leverage of the residualized treated group-time surface is $0.943$. The maintained sequence fixes this dimension and requires nonvanishing support shares and leverage tending to zero. The sample diagnostics show why that approximation is strained and the interval is wide. We therefore treat divorce as a finite-sample stress test with an inconclusive calibration, not as evidence that the published inference is invalid.

Taken together, the applications show why point envelopes, one-sided decisions, covariance rank, and orientation should be reported separately. Minimum wage fails under both additive and saturated lower bounds, and its estimated direction corroborates the concern. Brazil fails in every class with strong cohort support, low leverage, and full covariance rank. Castle-doctrine homicide passes for additive heterogeneity but lacks the covariance rank needed for a saturated certificate; its narrower family-adjusted outcome certificates reinforce the positive finding without broadening its scope. Divorce is unresolved despite large point values. A pass is uniform over orientations in the prespecified calibrated class under the maintained identification and asymptotic conditions; it is not uniform over those assumptions or over outcome-selected classes.

\section{Validation and practical implications}\label{sec:sim}

\subsection{Simulation evidence}

The simulations test separately the exact Gaussian map, the fixed-$T$ feasible bridge, and the new one-sided certificate. The baseline staggered design has $48$ units, $12$ periods, $\Gamma=1.54$, and an $11\%$ negative-weight share. Heterogeneity is aligned with the worst-case profile. Panel~A of Table~\ref{tab:validation} varies the resulting noncentrality. Panel~B holds $\eta_{CR}=2$ under AR(1) errors with $\rho=.5$ and increases the number of clusters.

\begin{table}[htbp]
\centering
\begin{threeparttable}
\caption{Finite-sample validation of the size map and feasible cluster scale}\label{tab:validation}
\small
\begin{tabular}{lrrr}
\toprule
index & Monte Carlo & oracle & Gaussian \\
\midrule
\multicolumn{4}{l}{\textit{Panel A: exact noncentrality calibration}}\\
$\eta=0.0$ & $0.050$ & -- & $0.050$ \\
$\eta=0.5$ & $0.079$ & -- & $0.079$ \\
$\eta=1.0$ & $0.170$ & -- & $0.170$ \\
$\eta=1.5$ & $0.322$ & -- & $0.323$ \\
$\eta=2.0$ & $0.516$ & -- & $0.516$ \\
\addlinespace
\multicolumn{4}{l}{\textit{Panel B: direct score calibration, $\eta_{CR}=2$}}\\
$G=48$  & $0.522$ & $0.514$ & $0.516$ \\
$G=120$ & $0.521$ & $0.519$ & $0.516$ \\
$G=300$ & $0.519$ & $0.518$ & $0.516$ \\
\bottomrule
\end{tabular}
\begin{tablenotes}[flushleft]\footnotesize
\item Notes: Entries are rejection frequencies for nominal-$5\%$ two-sided tests. Panel~A uses i.i.d. Gaussian errors. Panel~B uses AR(1) errors with $\rho=.5$; ``oracle'' divides by the population cluster-score scale, while the first numeric column uses~\eqref{eq:cr1}. Each row uses $200{,}000$ replications.
\end{tablenotes}
\end{threeparttable}
\end{table}

The first panel matches the noncentral prediction to three decimals. In this experiment the effect profile is aligned with the adverse design direction by construction, so the envelope is attained; the experiment validates the size map without asserting that an empirical profile has that orientation. Fourfold replication of the baseline panel halves absolute bias from $0.500$ to $0.250$ but changes rejection only from $0.322$ to $0.321$, verifying~\eqref{eq:bias}. Across independent, exchangeable, and AR(1) errors, oracle rejection also collapses onto the same curve in $\eta_{CR}$. In Panel~B the feasible-oracle gap is already eight-tenths of a percentage point at $G=48$ and becomes negligible as $G$ grows.

A separate experiment targets Theorem~\ref{thm:certificate}. It uses AR(1) errors with $\rho=.4$, $2{,}000$ replications, and $499$ multiplier draws. It reports the error rate and the probability of reaching each decision, rather than coverage alone. In a deliberately difficult three-cohort design with $G=240$, every wrong-decision rate is below $4\%$, but more than $94\%$ of samples are inconclusive. A favorable two-cohort design with $G=600$ certifies at $K=0$ in every replication and withholds in $74.4\%$ of replications at $K=1.0$, with no wrong decisions in either cell. The procedure therefore has practical power when exposure is moderate and group-time effects are estimated precisely. Full designs and additional covariance structures appear in Appendix~\ref{app:simulation}.

The empirical multiplier exercises test the relevant outcome-derived steps. Design $D$ and every $\Gamma_S$ are held fixed. A unit-plus-time-plus-group-time mean surface supplies residuals, but every draw and the covariance cross-product use the exact long-difference ATT$(g,t)$ cluster score map. The draws calibrate the lower norm radius and uncertainty in the signed contrast; the score cross-product supplies the covariance used by the upper equivalence test. Minimum wage's lower bounds remain above the adequacy threshold; castle's additive upper bound remains below it; and covariance-rank checks prevent saturated certificates in the two state panels. These are statements about the calibrated worst-case envelope, not estimates of realized rejection.

\subsection{Interpretation and use}\label{sec:discussion}

The minimum-wage application establishes that negative weights are not the operative condition and a comfortable sign-reversal diagnostic is not a uniform coverage guarantee, which is the direct confirmation of this article's main thesis. What matters for inference is the product of an outcome-free design exposure and heterogeneity measured in cluster-score noise units. In a large panel, a coefficient correction that is modest in levels can be many standard errors wide. Increasing the number of clusters shrinks that difference in levels along the local sequence, but it shrinks the conventional standard error at the same rate; therefore, the potential size distortion persists.

The diagnostic is intentionally conditional on the maintained DiD assumptions. It does not test parallel trends, select the scientifically preferred aggregation, or rescue a TWFE estimand that is substantively unattractive. Pre-trend and sensitivity analyses \citep{Roth,RambachanRoth} address violations of identification; heterogeneity-robust estimators address the target. The certificate answers the narrower question practitioners actually face after specifying a conventional TWFE regression: whether its test can be shown to remain size-controlled uniformly for a stated class and magnitude of heterogeneous effects.

The heterogeneity class must be chosen substantively and before inspecting the outcome pilot. Cohort, event-time, and additive profiles correspond to interpretable restrictions, while $S_{\mathrm{gt}}$ leaves the post-treatment cohort-time surface saturated. Reporting the full $\Gamma_S$ ladder shows which form of heterogeneity aligns with the design. When residualized treatment is constant within treated group-time cells, $S_{\mathrm{gt}}$ contains $w-u$, so within-cell effect variation cannot contribute to TWFE weighting bias. Covariates or observation patterns can break that closure and then $\Gamma_{\mathrm{gt}}<\Gamma$ must be reported. A pass for the additive class is narrower than a group-time pass. If $S$ is chosen from outcomes, simultaneous radii or sample splitting are required. The pilot and design statistic must always be matched to the same prespecified $S$.

Point envelope, one-sided decision bound, and direction answer different questions. $\Khat_S^{\mathrm{pt}}$ estimates the largest shift allowed by the calibrated magnitude; $\Khat^-_S$ tests for an unsafe envelope; $\Khat^+_S$ tests equivalence to the adequacy region; and $\widehat\eta_{\mathrm{dir},S}$ asks how the estimated profile aligns with the design. A small directional estimate cannot convert a failed uniform certificate into a pass. A large one strengthens the practical case for moving away from conventional TWFE, but still does not identify the true finite-sample rejection probability.

The direct cluster formulation is also practical. It accommodates arbitrary within-unit dependence under many clusters and does not require an AR(1) model. The separate $\sigma\sqrt\psi$ decomposition remains useful for interpretation but is not needed for implementation. With few clusters, CR2/CR3, a $t$ reference, or wild-cluster inference may be appropriate \citep{ImbensKolesar,CameronGelbachMiller,MNW}, but these procedures cannot be inserted into the Gaussian formula by swapping standard errors. The geometry $c_S\Gamma_S$ survives; the rejection map and critical threshold are procedure-specific and must be recalibrated for the chosen finite-$G$ test.

Uncertainty in the projected effect vector and in the score scale is part of the verdict. A lower bound above the threshold supports withholding certification; an upper bound below it supports a pass; otherwise the result is inconclusive. Minimum wage and Brazil illustrate the first outcome, castle-doctrine additive the second, and divorce the third. Reporting the directional interval alongside the one-sided decision prevents both a remote worst case and a noisy point orientation from being overstated.

\section{Conclusion}\label{sec:conclusion}

TWFE weighting bias becomes an inference problem when it is large relative to the cluster-score noise of the reported statistic. The exact noncentrality isolates that comparison as $c_S\Gamma_S/q_G$: $\Gamma_S$ records how the adoption design exposes the target to a specified form of heterogeneity, $c_S$ measures the magnitude of that heterogeneity, and $q_G$ puts the resulting bias in standard-error units. The direct score formulation makes the comparison feasible without separately estimating a residual scale and a serial-correlation correction. The reverse-triangle lower test and covariance-aware equivalence test remain regular at zero heterogeneity, while consistency of the score scale justifies ordinary plug-in under the bounded local experiment.

The implementation is intended to sit at the boundary between econometric theory and empirical practice. The adoption pattern determines $\Gamma_S$ before outcomes are examined, while the full verdict combines it with the coefficient, clustered score, and group-time estimates already produced in a modern DiD workflow; the accompanying software performs the audit directly.

The conclusion is deliberately asymmetric. A failed certificate directs the researcher to estimators robust to treatment-effect heterogeneity but neither selects the scientifically preferred target nor proves the published inference invalid. A pass certifies conventional TWFE inference only within the prespecified calibrated class and maintained DiD assumptions. Minimum wage cannot be certified even though all treated-cell weights are positive, which remains the paper's title result. Brazil shows that a failure can also be sharply identified in a large panel with strong support. Castle-doctrine homicide passes for additive heterogeneity but not for the saturated class, and its family-adjusted secondary outcomes pass only in narrower classes. Divorce shows why large point values need not be decisive when support is thin and the fixed-$T$, many-cluster approximation is strained. Point correction and conventional inference adequacy are separate empirical questions.

\section*{Acknowledgments}

I am deeply grateful to Sergey Alexeev (University of New South Wales) for generous discussions and invaluable suggestions and corrections. His comments materially strengthened the paper, sharpening both its theoretical claims and its empirical implementation. Any remaining errors are my own.

\section*{Data availability and reproducibility}

The minimum-wage application uses the exact public $2{,}284$-county replication data distributed by \citet{CS}; the Brazil application uses the public \citet{ChristensenGarfias2021} file distributed in the \citet{ChiuEtAl2026} reanalysis archive; the state applications use the public castle-doctrine panel distributed by \citet{ChengHoekstra} and the no-fault-divorce panel based on \citet{StevensonWolfers}. A replication archive containing the analysis-ready data, source provenance and checksums, audit scripts, fixed-design bootstrap code, and all reported results is available from the author. \softwareline

\section*{Disclosure statement}

The author has no conflicts of interest to declare.

\section*{Generative AI and AI-assisted technologies}

During preparation, the author used AI-assisted tools for supplementary consistency checks on mathematical arguments and code. The author verified and edited all content and takes full responsibility for the manuscript.

\clearpage
\appendix
\begin{center}\Large\bfseries Appendices\end{center}
\medskip
\section{Proofs}\label{app:proofs}

\subsection{Notation and elementary identities}

Let $M$ be the orthogonal projector that removes unit and time indicators, $\Dt=MD$, $\nw=\Dt'\Dt$, and $Q=M-\Dt\Dt'/\nw$. On the $N_1$ treated cells, $w_{it}=\Dt_{it}/\nw$ and $u_{it}=1/N_1$. Because $M$ is symmetric and idempotent,
\[
 \sum_{D=1}\Dt_{it}=D'MD=\Dt'\Dt=\nw,
\]
so $\onevec'w=1$ and $w-u\perp\onevec$.

\begin{lemma}[Cluster centering]\label{lem:center}
For every unit $i$, the residualized treatment path $d_i$ is cluster-centered:
\[
 d_i'\onevec=\sum_{t=1}^T\Dt_{it}=0.
\]
\end{lemma}

\begin{proof}
In a balanced panel,
\[
 \Dt_{it}=D_{it}-\bar D_{i\cdot}-\bar D_{\cdot t}+\bar D.
\]
Summing over $t$ gives $T\bar D_{i\cdot}-T\bar D_{i\cdot}-T\bar D+T\bar D=0$. More generally, the unit-indicator columns lie in the nuisance space, so orthogonality of $MD$ to those columns gives the same identity.
\end{proof}

\subsection{Exact numerator and studentization}

\begin{theorem}[Exact finite-sample statement]\label{thm:exact}
Suppose
\[
 Y_{it}=\alpha_i+\lambda_t+D_{it}\Delta_{it}+\eps_{it},
 \qquad \eps\mid D\sim N(0,\sigma^2I),
\]
and on treated cells $\Delta_{it}=\ATT+c s_{it}/\sqrt\nw$, with $\onevec's=0$ and $\|s\|^2=N_1$. Set
\[
 \eta=(c/\sigma)(w-u)'s,
 \qquad h=(c/\sqrt\nw)Q(D\circ s),
 \qquad \nu=\rank(Q).
\]
Then:
\begin{enumerate}[label=(\alph*)]
\item $\sqrt\nw(\hbeta-\ATT)/\sigma\sim N(\eta,1)$ exactly.
\item With $\hsigma^2=\|QY\|^2/\nu$, the numerator is independent of $\hsigma$, and
\[
 \frac{\nu\hsigma^2}{\sigma^2}\sim\chi^2_\nu(\|h\|^2/\sigma^2).
\]
Consequently the rejection probability of the usual studentized statistic, at any symmetric cutoff, is no larger than that of a noncentral $t_\nu(\eta)$. Equality holds when $Q(D\circ s)=0$.
\item If $N_1/\nw=O(1)$, $\nu\asymp n$, $c=O(1)$, and $\sigma$ is bounded above and away from zero, then $\hsigma^2\to_p\sigma^2$ and the studentized statistic converges to $N(\eta,1)$ along any sequence on which $\eta$ converges.
\end{enumerate}
\end{theorem}

\begin{proof}
Fixed effects vanish after premultiplication by $\Dt'$. Since $\Dt'D=\nw$ and the effect deviations live only on treated cells,
\begin{align*}
 \hbeta-\ATT
 &=\frac{c}{\nw^{3/2}}\Dt'(D\circ s)+\frac{\Dt'\eps}{\nw}\\
 &=\frac{c}{\sqrt\nw}w's+\frac{\Dt'\eps}{\nw}
 =\frac{c}{\sqrt\nw}(w-u)'s+\frac{\Dt'\eps}{\nw},
\end{align*}
where $u's=0$. The Gaussian error term has variance $\sigma^2/\nw$, proving (a).

The full regression residual is
\[
 QY=Q(D\circ\Delta)+Q\eps
 =\frac c{\sqrt\nw}Q(D\circ s)+Q\eps=h+Q\eps,
\]
because the common component $\ATT D$ lies in the regression space. Also
\[
 \Cov(Q\eps,\Dt'\eps)=\sigma^2Q\Dt=0.
\]
Joint normality therefore gives independence between the numerator and the residual vector. Since $Q$ is an orthogonal projector of rank $\nu$ and $h$ lies in its range, the usual noncentral quadratic-form result yields (b). A noncentral chi-square is first-order stochastically larger than its central counterpart. Conditional on the numerator, increasing the positive denominator weakly lowers the probability of crossing a symmetric cutoff, which gives the bound.

Finally, $Q$ is a contraction and $\|D\circ s\|^2=N_1$, so
\[
 \|h\|^2\le \frac{c^2N_1}{\nw}=O(1).
\]
Thus $\E\hsigma^2-\sigma^2=\|h\|^2/\nu=o(1)$ and
\[
 \Var(\hsigma^2)=\frac{2\sigma^4}{\nu^2}
 \left(\nu+2\frac{\|h\|^2}{\sigma^2}\right)=o(1).
\]
Part (c) follows from Chebyshev and Slutsky.
\end{proof}

\subsection{Worst cases and restricted profiles}

\begin{proposition}[Projection formula]\label{prop:projection}
Let $S\subseteq\R^{N_1}$ be linear, contain $\onevec$, and let $\widetilde S=S\cap\onevec^\perp$. If $\widetilde S\ne\{0\}$, then
\[
 \max_{s\in\widetilde S,\ \|s\|=\sqrt{N_1}}
 |(w-u)'s|
 =\sqrt{N_1}\|\Pi_{\widetilde S}(w-u)\|=: \Gamma_S.
\]
If $\widetilde S=\{0\}$, define $\Gamma_S=0$ and the only admissible centered profile has zero shift. If $S\subseteq S'$, then $\Gamma_S\le\Gamma_{S'}$; for $S=\R^{N_1}$, $\Gamma_S=\Gamma=\sqrt{N_1}\|w-u\|$.
\end{proposition}

\begin{proof}
For $s\in\widetilde S$, self-adjointness gives $(w-u)'s=[\Pi_{\widetilde S}(w-u)]'s$. Cauchy--Schwarz gives the upper bound, achieved by normalizing the projected vector whenever it is nonzero. Norms of orthogonal projections are monotone under nested subspaces. Since $w-u\perp\onevec$, unrestricted projection leaves it unchanged.
\end{proof}

For cohort and event-time partitions, projection replaces each element of $w-u$ by its within-cell-group mean. For the combined class, it is the least-squares fit on the union of cohort and event-time indicators, with the constant removed. Although the two sets of indicators are not orthogonal, the projection formula remains unchanged.

Let $S_{\mathrm{gt}}$ be the span of indicators for the observed post-treatment cohort-by-calendar-time cells.

\begin{proposition}[Group-time closure]\label{prop:gtclosure}
If residualized treatment is constant within every observed treated group-time cell,
\[
 w-u\in S_{\mathrm{gt}}\cap\onevec^\perp,
 \qquad \Gamma_{\mathrm{gt}}=\Gamma.
\]
Consequently only the group-time projection of an arbitrary treated-cell effect profile can contribute to TWFE weighting bias.
\end{proposition}

\begin{proof}
Constancy of $\Dt_{it}$ within $(g,t)$ makes $w$, and therefore $w-u$, an element of $S_{\mathrm{gt}}$. It is centered by the identity in Section~\ref{app:proofs}, so projection onto $S_{\mathrm{gt}}\cap\onevec^\perp$ leaves it unchanged and Proposition~\ref{prop:projection} gives $\Gamma_{\mathrm{gt}}=\Gamma$. For any profile $h$, self-adjointness gives $(w-u)'h=(w-u)'\Pi_{\widetilde S_{\mathrm{gt}}}h$; the orthogonal within-cell component drops out. A balanced unit-and-time-effects staggered design satisfies the premise because $D_{it}$, its unit mean, its time mean, and its grand mean depend on a treated observation only through $(g,t)$. Additional controls or unequal observation patterns need not satisfy it.
\end{proof}

\begin{proposition}[Envelope and directional shift]\label{prop:direction}
Let $h_S=\Pi_{\widetilde S}(\Delta-\bar\Delta\onevec)$ and define
\[
 c_S=\sqrt{\frac{\nw}{N_1}}\|h_S\|,
 \qquad
 r_S=\frac{\langle\Pi_{\widetilde S}(w-u),h_S\rangle}
 {\|\Pi_{\widetilde S}(w-u)\|\,\|h_S\|}
\]
when both norms are nonzero. Then $|r_S|\le1$ and the cluster-standardized mean shift generated by this particular profile is
\[
 \eta_{\mathrm{dir},S}
 =\frac{\sqrt\nw}{q_G}(w-u)'h_S
 =r_S\frac{c_S\Gamma_S}{q_G}.
\]
Thus $c_S\Gamma_S/q_G$ is a uniform envelope over orientations, while $r_S$ records the signed fraction attained by the profile.
\end{proposition}

\begin{proof}
Because $h_S\in\widetilde S$, self-adjointness gives
$ (w-u)'h_S=[\Pi_{\widetilde S}(w-u)]'h_S$. Multiplying the definitions of $c_S$ and $\Gamma_S$ yields
\[
 \frac{c_S\Gamma_S}{q_G}
 =\frac{\sqrt\nw}{q_G}
 \|\Pi_{\widetilde S}(w-u)\|\,\|h_S\|.
\]
The displayed identity follows after multiplication by the cosine $r_S$; Cauchy--Schwarz gives $|r_S|\le1$.
\end{proof}

With a group-time vector $\widehat\Delta$ and cell-incidence map $A$, the empirical signed contrast is
\[
 \widehat\eta_{\mathrm{dir},S}
 =\frac{\sqrt\nw}{\qhat}(w-u)'
 \Pi_{\widetilde S}(A\widehat\Delta-\overline{A\widehat\Delta}\onevec).
\]
It is linear in $\widehat\Delta$, so an unbiased group-time estimator requires no analogue of the quadratic trace subtraction used for the point estimate of $c_S^2$. Its uncertainty, including estimated orientation, is obtained by recomputing the contrast in the cluster multiplier bootstrap. Because the envelope uses a noise-corrected norm and the directional estimator is linear, the two finite-sample plug-ins need not obey the population factorization exactly.

\subsection{Projected-norm inference at the zero boundary}\label{app:normbound}

Let $P_S=\Pi_{\widetilde S}$ and define
\[
 \widehat m_S=\sqrt{\frac{\nw}{N_1}}P_SA\widehat\Delta,
 \qquad m_S=\sqrt{\frac{\nw}{N_1}}P_SA\Delta,
 \qquad c_S=\|m_S\|.
\]
For $\Omega=\Var(\widehat\Delta)$, the trace-debiased point calibration is
\begin{equation}\label{app:eq:dbpilot}
 (\chat_S^{\mathrm{db}})^2=\max\left\{0,
 \|\widehat m_S\|^2-
 \frac{\nw}{N_1}\tr(P_SA\widehat\Omega A'P_S)
 \right\}.
\end{equation}
The trace identity follows from $\E\|m_S+(\widehat m_S-m_S)\|^2=\|m_S\|^2+\tr\Var(\widehat m_S)$ when $\widehat\Delta$ is unbiased. The map $x\mapsto\sqrt{(x)_+}$ is nonregular at zero, so a percentile bootstrap of~\eqref{app:eq:dbpilot} is not used for certification.

Suppose instead that the group-time estimator has a fixed-dimensional cluster influence representation under fixed $T$,
\[
 \sqrt\nw(\widehat\Delta-\Delta)
 =\frac1{\sqrt G}\sum_{i=1}^G\xi_{Gi}+o_p(1),
\]
where the independent, mean-zero cluster influences satisfy a multivariate Lindeberg condition and have a limiting covariance that is nonsingular on the image of $P_SA$. Suppose the estimated influences obey
\[
 \frac1G\sum_i\|\widehat\xi_{Gi}-\xi_{Gi}\|^2=o_p(1).
\]
For independent multipliers $v_i$ with $\E v_i=0$, $\E v_i^2=1$, and a bounded $2+\delta$ moment, the standard multiplier CLT then gives, conditionally on the sample,
\begin{equation}\label{app:eq:bootconsistency}
 \sup_x\left|
 \Pp^*\!\left(\left\|\sqrt{\frac{\nw}{N_1}}P_SA
       (\widehat\Delta^*-\widehat\Delta)\right\|\le x\right)
 -\Pp\!\left(\|\widehat m_S-m_S\|\le x\right)
 \right|\to_p0.
\end{equation}
In the applications, each coordinate is the identified long difference
\[
 \widehat\Delta(g,t)=\overline{(Y_t-Y_{g-1})}_{G=g}
 -\overline{(Y_t-Y_{g-1})}_{C(g,t)},
\]
where $C(g,t)$ is the stated never-treated or not-yet-treated comparison pool. No anticipation and the corresponding parallel-trends restriction identify $\Delta(g,t)$. Every included cohort and comparison pool is required to have asymptotic share bounded away from zero, and the incidence map $A$ must cover every target treated cell. The exact cluster score places the two treated observations and two comparison observations in this display into their clusters with the same sample-mean weights. The saturated mean surface only supplies residuals for those scores; its regression coefficients are not substituted for $\widehat\Delta$. Rademacher multipliers are used in the applications. HC2 scaling based on leverage of that residual surface is the primary finite-sample adjustment; HC3 is reported as a sensitivity. Under vanishing group-time leverage both versions have the same limit.

Let $Q_S$ have orthonormal columns spanning the image of $P_S$. Define the projected vector in noncentrality units and its covariance by
\[
 \widehat x_S=\frac{\Gamma_S}{\qhat}Q_S'\widehat m_S,
 \qquad
 \widehat C_S=\widehat\Var(\widehat x_S).
\]
The target obeys $K_S=\|x_S\|$, where $x_S=(\Gamma_S/q_G)Q_S'm_S$. When $\widehat C_S$ has the prespecified rank $d_S=\dim(\widetilde S)$, form
\[
 W_S=\widehat x_S'\widehat C_S^{-1}\widehat x_S.
\]
For $W\sim\chi^2_{d_S}(\lambda)$, let $u_\gamma(w)$ be the nonnegative solution of
\[
 F_{d_S,u_\gamma(w)}(w)=\gamma,
\]
with $u_\gamma(w)=0$ when $F_{d_S,0}(w)\le\gamma$. This is the usual one-sided inversion of the noncentrality parameter.

\begin{theorem}[Regular one-sided decisions]\label{thm:normbound}
Let $\widehat r_{1-\gamma,S}$ be the conditional $(1-\gamma)$ quantile in~\eqref{app:eq:bootconsistency}, and assume that its limiting quantile is continuous. Suppose $c_S$ is bounded along the local sequence, $q_G$ is bounded away from zero, and $\qhat/q_G\to_p1$. Define
\begin{equation}\label{app:eq:lower}
 \Khat^-_S=\frac{\Gamma_S}{\qhat}
 (\|\widehat m_S\|-\widehat r_{1-\gamma,S})_+.
\end{equation}
Then
\[
 \liminf_G\Pp\{\Khat^-_S\le K_S\}\ge1-\gamma.
\]
Suppose in addition that $\widehat x_S\Rightarrow N(x_S,C_S)$ along every convergent local subsequence, $C_S$ is positive definite on the $d_S$-dimensional image of $P_S$, and $\widehat C_S\to_p C_S$. Define
\begin{equation}\label{app:eq:upper}
 \Khat^+_S=
 \sqrt{\lambda_{\max}(\widehat C_S)u_\gamma(W_S)}.
\end{equation}
Then
\[
 \liminf_G\Pp\{K_S\le\Khat^+_S\}\ge1-\gamma.
\]
Thus certification when $\Khat^+_S\le\eta^\dagger$ has asymptotic false-certification probability at most $\gamma$ whenever $K_S>\eta^\dagger$. Withholding certification when $\Khat^-_S>\eta^\dagger$ has the analogous error bound whenever $K_S<\eta^\dagger$. These are separate one-sided statements, not a joint $1-\gamma$ confidence interval. Both include $c_S=0$. If the estimated covariance lacks the prespecified rank, the upper procedure is unsupported and no certificate is issued.
\end{theorem}

\begin{proof}
For the lower statement, multiply the target by $\qhat/\Gamma_S$ and use
\[
 \left|\|\widehat m_S\|-c_S\frac{\qhat}{q_G}\right|
 \le \|\widehat m_S-m_S\|+c_S\left|\frac{\qhat}{q_G}-1\right|.
\]
The second term is $o_p(1)$ because $c_S$ is bounded. Bootstrap consistency, quantile continuity, and Slutsky's theorem therefore give
\[
 \Pp\left\{(\|\widehat m_S\|-\widehat r_{1-\gamma,S})_+
 \le c_S\frac{\qhat}{q_G}\right\}\ge1-\gamma+o(1),
\]
which is equivalent to the stated lower coverage. Relative consistency of $\qhat$ is enough; a deterministic denominator band is unnecessary in this bounded local experiment.

For the upper statement, Slutsky's theorem gives
\[
 W_S\Rightarrow\chi^2_{d_S}(\lambda_S),
 \qquad \lambda_S=x_S'C_S^{-1}x_S.
\]
For $\lambda_S>0$, monotonicity of the noncentral-$\chi^2$ cdf implies
\[
 \Pr_{\lambda_S}\{u_\gamma(W)<\lambda_S\}
 =\Pr_{\lambda_S}\{F_{d_S,\lambda_S}(W)<\gamma\}=\gamma;
\]
at $\lambda_S=0$, the upper bound covers identically. Hence $u_\gamma(W_S)$ is an asymptotic one-sided $1-\gamma$ upper bound for $\lambda_S$. Finally,
\[
 K_S^2=\|x_S\|^2
 \le\lambda_{\max}(C_S)x_S'C_S^{-1}x_S
 =\lambda_{\max}(C_S)\lambda_S.
\]
Eigenvalue consistency and another application of Slutsky's theorem prove the result along each convergent local subsequence, and therefore along the original sequence.
\end{proof}

The upper rule is a multivariate equivalence test: it asks whether the whole Euclidean norm lies below the adequacy margin. It follows the noncentrality-inversion logic in \citet{Lavergne2014} and the general equivalence-testing perspective in \citet{Romano2005}. The eigenvalue translation can remain conservative when $C_S$ is strongly anisotropic.

There is a more general confidence-set alternative. Let $a_G\downarrow0$ satisfy $\sqrt G a_G\to\infty$ when $\qhat/q_G-1=O_p(G^{-1/2})$, and retain the multiplier radius $\widehat r_{1-\gamma,S}$. Define
\begin{equation}\label{app:eq:ballalternative}
 \begin{split}
  \Khat^{-,\mathrm{ball}}_S
   &=(1-a_G)\frac{\Gamma_S}{\qhat}
     (\|\widehat m_S\|-\widehat r_{1-\gamma,S})_+,\\
  \Khat^{+,\mathrm{ball}}_S
   &=(1+a_G)\frac{\Gamma_S}{\qhat}
     (\|\widehat m_S\|+\widehat r_{1-\gamma,S}).
 \end{split}
\end{equation}
On the intersection of the multiplier confidence event and the denominator-band event, this interval contains $K_S$. Because the denominator event has probability tending to one, the interval has asymptotic joint coverage at least $1-\gamma$. This argument does not use boundedness of $c_S$ to discard denominator error, does not require a Gaussian projected vector, and can issue an upper statement without full covariance rank. Its scope is therefore broader than the decision-specific upper rule in Theorem~\ref{thm:normbound}.

The extra scope has a concrete power cost. The upper endpoint requires a full confidence ball around $\widehat x_S$ to fit inside the adequacy ball. If $C_S=\sigma_x^2I$ is known, its radius is $\sigma_x\sqrt{\chi^2_{d_S,1-\gamma}}$, so the procedure can never certify a margin below that radius, even at $x_S=0$. Noncentrality inversion instead uses the radial statistic directly and has positive power whenever the data are sufficiently close to zero; in the isotropic case it is the exact radial inversion. The ball construction is appropriate when $q_G$ is estimated by an unusual scheme for which an explicit rate band is more credible than ordinary plug-in, or when a nonsingular Gaussian projected covariance cannot be maintained. The paper therefore uses noncentrality inversion for certification under its bounded, fixed-dimensional Gaussian application experiment and keeps~\eqref{app:eq:ballalternative} as the more general option when those restrictions are not defensible. The narrower main rule gains power by using the maintained structure; it does not change the target or weaken the meaning of a certificate within that structure.

The probability statement is uniform over orientations in the fixed, prespecified subspace $S$ at the calibrated magnitude. It is conditional on the maintained identification and fixed-design assumptions and is not a simultaneous guarantee over outcome-selected subspaces. Selecting $S$ after examining $\widehat\Delta$ requires a simultaneous multiplier radius over the candidate family or sample splitting.

\begin{proposition}[Scale invariance along the local sequence]\label{prop:scale}
Replicate every unit $K$ times with the same adoption path and local profile, holding $c$ and $\sigma$ fixed. Then every $\Gamma_S$ and its worst-case noncentrality are unchanged, while the absolute weighting bias falls at rate $K^{-1/2}$.
\end{proposition}

\begin{proof}
Replication sends $N_1\mapsto KN_1$ and $\nw\mapsto K\nw$. Each replicated treated cell receives weight $w_{it}/K$ and uniform weight $u_{it}/K$. Hence the squared norm of the replicated imbalance is $K\|(w-u)/K\|^2=\|w-u\|^2/K$, which exactly offsets the multiplication of $N_1$ in $\Gamma_S^2$. From Theorem~\ref{thm:exact},
\[
 |\E\hbeta-\ATT|=\frac{|\eta|\sigma}{\sqrt\nw},
\]
and the denominator grows by $\sqrt K$.
\end{proof}

\subsection{Oracle cluster normalization}

Let clusters be independent with $\E(\eps_i\mid D)=0$ and $\Cov(\eps_i\mid D)=\sigma^2R_i$. Define
\[
 \psi_G=\frac{\sum_i d_i'R_i d_i}{\nw},
 \qquad q_G^2=\sigma^2\psi_G.
\]

\begin{proposition}[Oracle cluster mean shift]\label{prop:oracle}
Under joint Gaussianity and $q_G>0$,
\[
 \frac{\hbeta-\ATT}{q_G/\sqrt\nw}
 \sim N\left(\frac c{q_G}(w-u)'s,1\right).
\]
The worst-case mean shift over $S$ is $c\Gamma_S/q_G=(c/\sigma)\Gamma_S/\sqrt{\psi_G}$.
\end{proposition}

\begin{proof}
The deterministic weighting bias is unchanged. Independence across clusters gives
\[
 \Var(\Dt'\eps)=\sum_i\Var(d_i'\eps_i)=\nw q_G^2.
\]
The assertion follows by Gaussianity and the projection argument above.
\end{proof}

Two reference covariance structures show why the direction of clustering cannot be assumed. If $R_i=(1-\rho_c)I+\rho_c\onevec\onevec'$, Lemma~\ref{lem:center} gives $\psi_G=1-\rho_c$, so a positive common component is annihilated by the unit effect and clustering raises the standardized exposure. Under AR(1), $d_i'R_i d_i$ can place $\psi_G$ on either side of one; it must be computed from the realized paths.

\subsection{Proof of the \texorpdfstring{fixed-$T$}{fixed-T} feasible result}\label{app:fixedT}

We now justify the explicitly defined cluster-score sandwich. Let
\[
 S_i=d_i'\eps_i,\qquad
 q_G^2=\nw^{-1}\sum_i\Var(S_i\mid D),
\]
and let $\widehat S_i=d_i'\widehat\eps_i$ be the cluster score from the full TWFE residual. Conditional mean zero makes $\Var(S_i\mid D)=\E(S_i^2\mid D)$. The cluster-only corrected scale is
\[
 \qhat^2=\frac{G}{G-1}\frac1\nw\sum_i\widehat S_i^2.
\]

\begin{theorem}[Feasible score sandwich with fixed $T$]\label{thm:fixedT}
Suppose $T$ is fixed and clusters are independent with $\E(\eps_i\mid D)=0$. For some $\delta>0$, assume
\[
 \sup_i\E\|\eps_i\|^{4+\delta}<\infty,
 \qquad \nw/G\asymp1,
 \qquad q_G^2\asymp1,
 \qquad \frac{\max_i\|d_i\|^2}{\nw}\to0,
\]
and
\[
 \frac{\sum_i\E|S_i|^{2+\delta}}
 {(\nw q_G^2)^{1+\delta/2}}\to0.
\]
With a fixed number of common time indicators and bounded local scale $c$,
\[
 \qhat^2-q_G^2=O_p(G^{-1/2}),
 \qquad \qhat/q_G-1=O_p(G^{-1/2}),
\]
and the feasible score-sandwich statistic centered on $\ATT$ converges to a unit-variance Gaussian with mean $c(w-u)'s/q_G$, whenever that mean converges.
\end{theorem}

\begin{proof}
The Lyapunov condition gives
\begin{equation}\label{eq:clt}
 \frac{\sum_iS_i}{\sqrt{\nw q_G^2}}\Rightarrow N(0,1).
\end{equation}
It remains to show that the residual score meat estimates $\sum_i\E S_i^2$.

Write the error after removing unit and time indicators as
\[
 (M\eps)_{it}=\eps_{it}-\bar\eps_{i\cdot}-\bar\eps_{\cdot t}+\bar\eps.
\]
By Lemma~\ref{lem:center}, both terms constant within $i$ vanish after premultiplication by $d_i'$. Hence
\[
 d_i'(M\eps)_i=S_i-a_i,
 \qquad a_i=\sum_t d_{it}\bar\eps_{\cdot t}.
\]
Because $T$ is fixed and each time mean averages $G$ independent clusters, $\|\bar\eps_{\cdot}\|=O_p(G^{-1/2})$. Therefore
\[
 \sum_i a_i^2
 \le \sum_i\|d_i\|^2\|\bar\eps_{\cdot}\|^2
 =\nw O_p(G^{-1})=O_p(1),
\]
so the normalized sum is $O_p(G^{-1})$.

Estimating the treatment slope subtracts $b_i=(d_i'd_i)(\hbeta-\beta_D)$ from cluster $i$, where $\beta_D$ denotes the conditional weighted mean under the local drift. Since $\hbeta-\beta_D=O_p(\nw^{-1/2})$,
\[
 \frac1\nw\sum_i b_i^2
 \le (\hbeta-\beta_D)^2\max_i\|d_i\|^2=O_p(G^{-1}),
\]
where fixed $T$ and binary treatment imply $\max_i\|d_i\|=O(1)$.
For the residualized local-heterogeneity vector
\[
 h=\frac c{\sqrt\nw}Q(D\circ s),
\]
no cellwise bound is needed. Since $Q$ is a contraction, fixed $T$, $\nw/G\asymp1$, and $\|s\|^2=N_1$ imply
\[
 \|h\|^2\le \frac{c^2N_1}{\nw}=O(1).
\]
Its cluster-score contribution $\ell_i=d_i'h_i$ therefore satisfies the aggregate bound
\[
 \frac1\nw\sum_i\ell_i^2
 \le\frac{\max_i\|d_i\|^2}{\nw}\|h\|^2=O(G^{-1}).
\]
Combining these components gives
\begin{equation}\label{eq:scoreapprox}
 \frac1\nw\sum_i(\widehat S_i-S_i)^2=O_p(G^{-1}).
\end{equation}
By Cauchy--Schwarz, the normalized cross term is $O_p(G^{-1/2})$. This follows from~\eqref{eq:scoreapprox} and $\nw^{-1}\sum_iS_i^2=O_p(1)$.

Finally, the fourth-moment bound, $T$ fixed, and $\nw\asymp G$ yield the triangular-array rate
\[
 \frac1\nw\sum_i(S_i^2-\E S_i^2)=O_p(G^{-1/2}).
\]
Since $G/(G-1)=1+O(G^{-1})$, the preceding bounds give $\qhat^2-q_G^2=O_p(G^{-1/2})$. Because $q_G$ is bounded away from zero, the relative rate follows by the delta method. Combine this result,~\eqref{eq:clt}, and the deterministic mean shift from Theorem~\ref{thm:exact} by Slutsky.
\end{proof}

The theorem is deliberately specialized. \citet{AnatolyevNg2026} allow many covariates, unbalanced and asymptotically unbounded clusters, and develop variance estimators for that broader problem. Here fixed $T$ and cluster-centered within paths make the unit-effect contribution exactly zero, permitting the score meat under the stated leverage condition. Our finite-sample convention multiplies only by $G/(G-1)$; it does not use an additional $(n-1)/(n-K)$ factor.

\subsection{Size inversion}

Let $z=z_{1-\alpha/2}$ and
\[
 p(\eta)=\Phi(-z-\eta)+1-\Phi(z-\eta).
\]
Using $\Phi(-x)=1-\Phi(x)$ shows $p(\eta)=p(-\eta)$. Moreover
\[
 p'(\eta)=\phi(z-\eta)-\phi(z+\eta)>0\quad(\eta>0),
\]
and $p''(0)=2z\phi(z)$. Thus $p$ is strictly increasing in $|\eta|$ and
\[
 p(\eta)=\alpha+z\phi(z)\eta^2+O(\eta^4).
\]
The unique positive solution to $p(\eta)=\alpha+\delta$ is the adequacy threshold. At $\alpha=\delta=.05$, numerical inversion gives $\eta^\dagger=0.652358$; the quadratic approximation is $0.660674$. A linear-in-$|\eta|$ approximation would be incorrect because the two-sided size is even.

\section{Monte Carlo details}\label{app:simulation}

All experiments condition on the adoption pattern. The baseline design contains $48$ units and $12$ periods, with three equal adoption cohorts at periods $3$, $7$, and $11$ and no never-treated units. It has $\Gamma=1.54$ and an $11\%$ negative-weight share. Heterogeneity is aligned with the worst-case profile. Each row uses $200{,}000$ draws unless otherwise noted.

\begin{table}[htbp]\centering
\caption{Exact noncentrality calibration}\label{tab:mc}
\begin{tabular}{rrrr}
\toprule
$\eta$ & Monte Carlo & prediction & difference \\
\midrule
$0.0$ & $0.050$ & $0.050$ & $0.000$ \\
$0.5$ & $0.079$ & $0.079$ & $<0.001$ \\
$1.0$ & $0.170$ & $0.170$ & $<0.001$ \\
$1.5$ & $0.322$ & $0.323$ & $-0.001$ \\
$2.0$ & $0.516$ & $0.516$ & $<0.001$ \\
\bottomrule
\end{tabular}
\end{table}

\begin{table}[htbp]\centering
\caption{Scale invariance along the local sequence}\label{tab:scale}
\begin{tabular}{lrrrr}
\toprule
panel & $\nw$ & $\Gamma$ & absolute bias & rejection \\
\midrule
$G=48$  & $42.7$  & $1.54$ & $0.500$ & $0.322$ \\
$G=192$ & $170.7$ & $1.54$ & $0.250$ & $0.321$ \\
\bottomrule
\end{tabular}
\end{table}

\begin{table}[htbp]\centering
\caption{Oracle cluster normalization}\label{tab:cluster}
\begin{tabular}{lrrrrr}
\toprule
within-unit model & parameter & $\psi$ & $\Gamma/\sqrt\psi$ & MC & prediction \\
\midrule
i.i.d. & -- & $1.00$ & $1.54$ & $0.516$ & $0.516$ \\
exchangeable & $\rho_c=.5$ & $0.50$ & $2.18$ & $0.519$ & $0.516$ \\
AR(1) & $\rho=.5$ & $1.48$ & $1.27$ & $0.514$ & $0.516$ \\
AR(1) & $\rho=.8$ & $1.16$ & $1.43$ & $0.513$ & $0.516$ \\
\bottomrule
\end{tabular}
\end{table}

For the feasible bridge, the target is $\eta_{CR}=2$ under AR(1) errors with $\rho=.5$. At $G=48$, feasible score-sandwich rejection is $0.522$, oracle rejection $0.514$, and the Gaussian prediction $0.516$. At $G=120$, feasible and oracle rates are $0.521$ and $0.519$; at $G=300$, $0.519$ and $0.518$. The shrinking feasible-oracle gap matches the fixed-$T$ convergence in Section~\ref{app:fixedT}.

\begin{table}[htbp]\centering
\begin{threeparttable}
\caption{One-sided decision performance}\label{tab:normmc}
\begin{tabular}{lrrrrr}
\toprule
design & population $K$ & certify & withhold & inconclusive & wrong decision \\
\midrule
hard & $0.00$ & $0.0545$ & $0.0040$ & $0.9415$ & $0.0040$ \\
hard & $0.60$ & $0.0360$ & $0.0085$ & $0.9555$ & $0.0085$ \\
hard & $0.70$ & $0.0390$ & $0.0085$ & $0.9525$ & $0.0390$ \\
favorable & $0.00$ & $1.0000$ & $0.0000$ & $0.0000$ & $0.0000$ \\
favorable & $0.70$ & $0.0000$ & $0.0030$ & $0.9970$ & $0.0000$ \\
favorable & $1.00$ & $0.0000$ & $0.7435$ & $0.2565$ & $0.0000$ \\
\bottomrule
\end{tabular}
\begin{tablenotes}[flushleft]\footnotesize
\item Notes: The hard design has $G=240$, $T=8$, three adoption cohorts and a never-treated group. The favorable design has $G=600$, $T=8$, two late cohorts and 480 never-treated clusters. Errors are AR(1) with $\rho=.4$; each cell uses $2{,}000$ Monte Carlo replications and $499$ Rademacher draws. The threshold is $0.652358$. ``Wrong decision'' means certification when $K$ is above the threshold or withholding when $K$ is below it. The upper decision uses covariance-aware noncentral-$\chi^2$ inversion; the lower decision uses the multiplier norm radius. The experiment uses oracle group-time influence scores and a feasible score sandwich for $q_G$ to isolate Theorem~\ref{thm:normbound}. Table~\ref{tab:cluster} separately validates replacement of error scores by regression-residual scores.
\end{tablenotes}
\end{threeparttable}
\end{table}

The hard design deliberately puts a high-dimensional group-time vector against a moderate number of clusters. The error rates remain below the nominal $\gamma=5\%$ bound, but most samples are inconclusive. The favorable design provides the complementary power check: at homogeneity it certifies in every replication; at $K=1.00$ it withholds in $74.35\%$ and is otherwise inconclusive. The near-boundary $K=0.70$ cell is almost always inconclusive, as a level-$5\%$ decision procedure should be close to its boundary. Reporting decision frequencies makes the method's useful and weak regions visible instead of using coverage alone.

\section{Application construction}\label{app:construction}

\subsection{Common audit}

For each panel we perform the same operations:
\begin{enumerate}[label=\arabic*.]
\item Construct the post-adoption indicator, residualize it on the specification's fixed effects, and compute $w$, $\Gamma$, and the cohort/event/additive/group-time projections.
\item Form group-time effects using the comparison group specified by the original application. Let $A$ duplicate each group-time estimate over the treated cells it represents.
\item Fit a saturated unit, time, and treated group-time mean surface to obtain residuals. Construct the exact cluster influence scores of the long-difference ATT$(g,t)$ estimator from those residuals. A code assertion verifies that summing these scores reproduces every residualized ATT$(g,t)$ contrast. County-, state-, or municipality-level Rademacher multipliers applied to these estimator scores generate $999$ fixed-design draws for the headline applications and $9{,}999$ draws for the extended castle-outcome and Brazil audit. HC2 scales residuals for the primary norm radius; HC3 is a sensitivity.
\item Use the exact cross-product of the ATT$(g,t)$ score matrix for $\widehat\Omega$ in the projected trace-debiased point calibration. This removes avoidable simulation noise from the point quantity. Use the $(1-\gamma)$ quantile of the projected centered multiplier increment for the lower bound and the same score cross-product for the upper equivalence test.
\item Map the point calibration through Gaussian size. Compare the separate one-sided bounds $\Khat^-_S$ and $\Khat^+_S$ with the adequacy threshold, and issue no upper certificate if the covariance lacks the prespecified rank. Separately compute the signed directional contrast from the estimated group-time profile.
\end{enumerate}

Holding treatment fixed is deliberate: $\Gamma_S$ is a known feature of the realized adoption design, while the bootstrap quantifies uncertainty in the outcome-derived calibration. A cluster pairs bootstrap, which lets the adoption composition vary, is reported as a sensitivity for the state applications.

\subsection{Minimum wage}\label{app:minwage}

The data file \texttt{min\_wage\_CS.rds} is the authors' exact public analysis file: $2{,}284$ counties observed annually from 2001 through 2007. The 2004, 2006, and 2007 adoption groups contain 100, 223, and 584 counties; 1,377 counties remain at the federal minimum throughout the sample. The outcome is log teen employment. We use the paper's unconditional design, never-treated controls, and no anticipation. The TWFE coefficient is $-0.036609$, reproducing the published $-0.037$ after rounding. Cohort sizes and exposure lengths imply $400+446+584=1{,}430$ treated county-year cells, all represented by the seven post-treatment group-time effects. Weighting those effects by their numbers of treated county-year observations gives the target-matched robust aggregate $-0.051581$; the published $-0.039$ group aggregate uses different cohort weights.

\begin{table}[htbp]\centering
\caption{Minimum-wage audit: exact published sample}\label{tab:minwage}
\small
\begin{tabular}{lr@{\qquad}lr}
\toprule
quantity & value & quantity & value \\
\midrule
counties & $2{,}284$ & treated cells & $1{,}430$ \\
$\widehat\beta_{TWFE}$ & $-0.03661$ & target-matched robust ATT & $-0.05158$ \\
$\nw$-residual scale $\widehat\sigma$ & $0.14319$ & negative-weight share & $0.000$ \\
$\Gamma=\Gamma_{\mathrm{gt}}$ & $0.26964$ & $\Gamma_{\mathrm{c+e}}$ & $0.25728$ \\
$\Gamma_{\mathrm{coh}}$ & $0.16600$ & $\Gamma_{\mathrm{evt}}$ & $0.25091$ \\
$\bar\sigma_0=|\widehat\beta|/\Gamma$ & $0.13577$ & directional alignment $\widehat r$ & $0.92476$ \\
$\chat^{\mathrm{db}}_{\mathrm{gt}}/\widehat\sigma$ & $6.46015$ & direct $\widehat\psi$ & $1.38489$ \\
$\Khat^{\mathrm{pt}}_{\mathrm{gt}}$ & $1.48021$ & point size envelope & $0.31599$ \\
$\Khat^-_{\mathrm{gt}}$ (HC2, $95\%$) & $0.79341$ & lower-bound size & $0.12464$ \\
$\Khat^-_{\mathrm{c+e}}$ (HC2, $95\%$) & $0.75629$ & lower-bound size & $0.11766$ \\
$\Khat^-_{\mathrm{gt}}$ (HC3, $95\%$) & $0.79215$ & threshold $\eta^\dagger$ & $0.65236$ \\
$\widehat\eta_{\mathrm{dir}}$ & $1.37026$ & directional size & $0.27813$ \\
directional wild median & $0.23461$ & directional wild $95\%$ & $[0.11780,0.40496]$ \\
\bottomrule
\end{tabular}
\end{table}

An AR(1) working model gives $\widehat\rho=.2657$ and $\widehat\psi_{AR1}=1.0915$; the direct score scale requires no such working correlation model. The saturated point envelope is close to its additive counterpart ($31.6\%$ versus $29.3\%$). The reverse-triangle lower bounds are $0.793$ for the saturated class and $0.756$ for the additive class, both above the $0.652$ threshold. HC3 changes the saturated value to $0.792$. Concern is also not supported only by an adversarial rotation after estimating magnitude: the additive profile's alignment is $0.925$, and its directional interval remains above the $10\%$ adequacy ceiling. These calculations justify withholding a certificate under both reported classes; they do not estimate rejection under a known true effect profile.

\subsection{Brazilian property taxation}\label{app:brazil}

\citet{ChristensenGarfias2021} estimate the fiscal effect of municipal cadastre updates in Brazil. We use the public Christensen--Garfias file in the reanalysis archive of \citet{ChiuEtAl2026} and follow the unit-and-year fixed-effects specification for log property-tax revenue in Table~2, column~1. The raw file has $64{,}698$ rows, $5{,}568$ municipalities, and 12 years from 2004 through 2015. Requiring a complete 12-year path leaves $3{,}810$ municipalities; excluding 970 municipalities already treated in 2004 leaves $2{,}840$ with an observed untreated baseline. This balanced-subset construction changes the estimation sample and target from the full published analysis, so the exercise is reported as a reanalysis of the specification rather than an exact coefficient reproduction.

The balanced-sample TWFE coefficient is $0.059346$, with municipality-clustered standard error $0.026788$. The equally treated-cell-weighted long-difference aggregate is $0.028459$. Of the treated-cell TWFE weights, $17.34\%$ are negative. The design statistics are
\[
 (\Gamma_{\mathrm{coh}},\Gamma_{\mathrm{evt}},
  \Gamma_{\mathrm{c+e}},\Gamma_{\mathrm{gt}},\Gamma)
 =(0.46277,1.02524,1.06482,1.13808,1.13808).
\]

\begin{table}[htbp]
\centering
\caption{Brazil property-tax audit on the balanced subset}\label{tab:brazil}
\small
\begin{threeparttable}
\begin{tabular}{lrrrrrr}
\toprule
class & $\Gamma_S$ & $\Khat_S^{\mathrm{pt}}$ & $\Khat^-_{S,HC2}$ & $\Khat^-_{S,HC3}$ & HC2 rank & decision \\
\midrule
cohort & 0.463 & 3.053 & 1.040 & 1.036 & $10/10$ & withhold \\
event time & 1.025 & 6.021 & 2.441 & 2.434 & $10/10$ & withhold \\
additive & 1.065 & 8.620 & 3.342 & 3.334 & $20/20$ & withhold \\
group time & 1.138 & 9.688 & 3.223 & 3.210 & $65/65$ & withhold \\
\bottomrule
\end{tabular}
\begin{tablenotes}[flushleft]\footnotesize
\item Notes: Each lower endpoint is a separate one-sided $95\%$ reverse-triangle bound based on $9{,}999$ municipality-level Rademacher draws. HC3 has the same displayed full ranks as HC2. The decision threshold is $0.652358$. The smallest treated cohort has 88 municipalities, the smallest comparison pool has 99, and maximum group-time leverage is $0.01009$.
\end{tablenotes}
\end{threeparttable}
\end{table}

All four lower bounds exceed the threshold under both leverage corrections. The cohort class is the least adverse and its HC2 lower endpoint still exceeds the threshold by $0.388$. Full projected covariance rank, low leverage, and nontrivial cohort and comparison support distinguish this result from the state-panel rank failures. These diagnostics support the precision of the calibration under the maintained fixed-$T$, many-municipality experiment. They do not establish the identifying assumptions or imply that the realized treatment-effect profile attains the worst-case orientation.

\subsection{Castle doctrine}\label{app:castle}

The \citet{ChengHoekstra} panel contains $50$ states from 2000 through 2010; 21 adopt and 29 never adopt. The outcome is log homicide. The TWFE coefficient is $0.081812$. The adoption statistics are
\[
 (\Gamma_{\mathrm{coh}},\Gamma_{\mathrm{evt}},
 \Gamma_{\mathrm{c+e}},\Gamma_{\mathrm{gt}},\Gamma)
 =(0.14215,0.16780,0.19818,0.21142,0.21142),
\]
with no negative treated-cell weights. The direct residual-score ratio is $\widehat\psi=3.3721$; the older AR(1) decomposition gives $\widehat\rho=.2264$ and $\widehat\psi_{AR1}=1.3358$. After subtracting the exact fitted-score covariance, the cohort, event-time, and additive pilots are at zero. The saturated pilot is $\chat^{\mathrm{db}}_{\mathrm{gt}}/\widehat\sigma=0.74787$, yielding $\Khat^{\mathrm{pt}}_{\mathrm{gt}}=0.08610$ and a $5.08\%$ point envelope.

Covariance-aware noncentral-$\chi^2$ inversion gives the additive class an HC2 upper endpoint of $0.23579$, below the $0.65236$ threshold; HC3 gives $0.22456$. The saturated score covariance has rank $18$, while the centered saturated subspace requires rank $19$, so the algorithm issues no saturated certificate. This illustrates a practical consequence of thin support: with few clusters, a structured class can be informative when a saturated surface cannot support inference in every direction. The smallest adoption cohort contains one state and maximum treated group-time leverage is $0.89091$, so the pass remains conditional on the maintained many-cluster approximation. The sign-reversal dispersion is $\bar\sigma_0=0.3870$. For the additive profile, $\widehat\eta_{\mathrm{dir}}=-0.0134$, corresponding to $5.0\%$ directional size, and its descriptive wild interval is $[5.0,5.5]\%$.

Homicide in the additive class remains the headline certificate. Table~\ref{tab:castlefamily} reports the secondary outcome-class pairs that also survive adjustment for searching the complete recorded family of 20 outcomes and four heterogeneity classes. The cohort and event-time spaces are fixed from adoption timing before the outcome pilot is inspected. The familywise column uses the Bonferroni level $0.05/80$ inside the same noncentral-$\chi^2$ inversion.

\begin{table}[htbp]
\centering
\caption{Castle-doctrine certificates across outcomes and classes}\label{tab:castlefamily}
\small
\begin{threeparttable}
\begin{tabular}{llrrrrc}
\toprule
outcome & class & \multicolumn{2}{c}{ordinary $95\%$ upper} & \multicolumn{2}{c}{family-80 upper} & family result \\
\cmidrule(lr){3-4}\cmidrule(lr){5-6}
& & HC2 & HC3 & HC2 & HC3 & \\
\midrule
homicide & additive & 0.236 & 0.225 & 0.457 & 0.468 & certify \\
burglary & cohort & 0.218 & 0.222 & 0.439 & 0.451 & certify \\
burglary & event time & 0.217 & 0.211 & 0.381 & 0.379 & certify \\
robbery & cohort & 0.226 & 0.194 & 0.448 & 0.440 & certify \\
robbery & event time & 0.443 & 0.435 & 0.584 & 0.581 & certify \\
assault & event time & 0.355 & 0.380 & 0.434 & 0.467 & certify \\
\bottomrule
\end{tabular}
\begin{tablenotes}[flushleft]\footnotesize
\item Notes: Ordinary columns invert at one-sided level $0.05$; family columns invert at $0.05/80$. Every family endpoint shown is below $0.652358$ under both HC2 and HC3. Assault-cohort passes pointwise ($0.592$ HC2; $0.622$ HC3) but not after family adjustment ($0.789$; $0.844$), so it is not counted as a family-adjusted certificate. Additive upper endpoints for burglary, robbery, and assault exceed the threshold. Every saturated outcome covariance has rank $18/19$, so no saturated upper certificate is issued.
\end{tablenotes}
\end{threeparttable}
\end{table}

The secondary results sharpen the interpretation of the split verdict. At $G=50$, several economically simple cohort or event-time profiles are estimable tightly enough to certify even after a broad family adjustment. Allowing additive heterogeneity can already be too ambitious for the non-homicide outcomes, and the saturated class is unsupported in one direction for every outcome. The positive evidence is therefore real but explicitly class-specific.

\subsection{No-fault divorce}\label{app:divorce}

The no-fault-divorce panel uses the female-suicide outcome in \citet{StevensonWolfers} from 1964 through 1990. Two states treated before the sample begins lack an in-sample untreated baseline and are excluded from the group-time construction. The resulting balanced panel has $49$ states, 20 never-treated states, and 522 treated state-year cells. The conventional TWFE coefficient is $-0.015783$; $1.15\%$ of treated-cell weights are negative. Its design statistics are
\[
 (\Gamma_{\mathrm{coh}},\Gamma_{\mathrm{evt}},
 \Gamma_{\mathrm{c+e}},\Gamma_{\mathrm{gt}},\Gamma)
 =(0.38096,0.46809,0.56236,0.64326,0.64326),
\]
and $\bar\sigma_0=0.02454$.

The trace-debiased saturated pilot is $6.59942$ in residual-noise units. The score sandwich gives $\widehat\psi=2.91718$, $\Khat^{\mathrm{pt}}_{\mathrm{gt}}=2.48547$, and a $70.0\%$ point envelope. Its reverse-triangle lower bound is $0.08342$, below the adequacy threshold. The saturated score covariance has rank $49$ against the $167$ directions required for a saturated certificate, so no upper bound is issued; the additive equivalence bound also lies far above the threshold. The estimated additive profile has alignment $0.5678$ and directional shift $1.6292$, corresponding to $37.1\%$ directional size; its descriptive interval is $[5.0,47.6]\%$. Both calculations are therefore inconclusive once sampling uncertainty is included.

This panel also provides only 49 independent clusters with 27 observations per cluster. Its pilot has 168 identified group-time coordinates and centered saturated rank 167. Several adoption cohorts contain one state, and maximum leverage of the residualized treated group-time surface is $0.943$. The maintained sequence fixes the pilot dimension, requires treated and comparison shares bounded away from zero, and requires leverage to vanish. These diagnostics make divorce the weakest approximation among the four applications and account for the rank failure. The calculations are reported as a finite-sample stress test, not as a claim that the published inference is invalid.

\section{Extensions and implementation notes}\label{app:extensions}

\subsection{Covariates}

If the reported TWFE model includes observed controls, replace $M$ by the residual maker for unit effects, time effects, and those controls. All weight and projection formulas continue to use $\Dt=MD$. The equality $\Gamma_{\mathrm{gt}}=\Gamma$ must then be checked: it holds only if the new $\Dt$ remains constant within treated group-time cells. Otherwise $\Gamma_{\mathrm{gt}}<\Gamma$ is merely a restricted exposure and within-cell heterogeneity can matter. The direct cluster theorem also requires the nuisance projection error in each treatment score to be negligible; fixed-dimensional common controls satisfy the same argument as time effects. With many controls, a cluster-and-many-covariate robust estimator satisfying the conditions in \citet{AnatolyevNg2026} can estimate $q/\sqrt\nw$ for the same asymptotic Gaussian test.

\subsection{Unbalanced panels}

The design statistic itself needs only an orthogonal residualization of $D$ on the fixed-effect design, so it is unchanged conceptually in an unbalanced panel. The exact cluster-centering identity continues to hold because the unit indicator is in the nuisance space: the sum of $\Dt_{it}$ over the observations belonging to unit $i$ is zero. Unequal observation patterns can make $\Dt$ vary among units in the same treated group-time cell, so group-time closure and $\Gamma_{\mathrm{gt}}=\Gamma$ must be verified rather than assumed. For asymptotically unbounded cluster size, the fixed-$T$ proof no longer applies directly; the cluster-size and auxiliary-projection conditions of \citet{AnatolyevNg2026} are the appropriate route.

\subsection{Few clusters}

Consistency of the score sandwich is a many-cluster statement. With few clusters, practitioners may prefer CR2/CR3, a $t$ reference distribution with effective degrees of freedom, or wild-cluster inference \citep{ImbensKolesar,CameronGelbachMiller,MNW}. Only the geometric numerator $c_S\Gamma_S$ transfers automatically. A $t$ or bootstrap test has a procedure-specific rejection map, so its standard error cannot simply replace $\widehat{\mathrm{SE}}_C$ in the Gaussian threshold. Extending the certificate requires inverting the actual finite-$G$ procedure.

\subsection{Replication archive}

The archive is organized as follows:
\begin{itemize}
\item \texttt{data/}: the minimum-wage, castle-doctrine, divorce, and Brazil analysis files, source URLs, licences where supplied, and SHA-256 checksums;
\item \texttt{code/screen\_minimum\_wage.R}: dependency-free exact-sample minimum-wage audit and $999$-draw fixed-design county wild bootstrap;
\item \path{code/gate2_crgamma_audit.py} and
\path{code/gate2_restricted_gamma.py}: the state-application audit and matched projections;
\item \path{code/gate2_bootstrap.py}: covariance correction, lower norm bounds, covariance-aware equivalence tests, rank diagnostics, and wild/pairs bootstraps;
\item \path{code/gate2_screen_core.py} and \path{code/gate2_extended_applications.py}: compressed cohort-time implementation and the reproducible castle family/Brazil audit, including the family-80 adjustment;
\item \path{simulations/gate2_norm_bounds_sim.py}: boundary and decision-error experiment in Table~\ref{tab:normmc};
\item \texttt{results/}: machine-readable outputs for every application table and simulation;
\item \texttt{simulations/}: scripts and frozen outputs for Tables~\ref{tab:mc}--\ref{tab:normmc}.
\end{itemize}

{\small
\bibliography{paper_c_bib}
}

\end{document}